\documentclass[12pt]{JHEP}
\title{A road map to solar neutrino fluxes, neutrino oscillation parameters, and tests for new physics}

\author{John N. Bahcall\\
  School of Natural Sciences, Institute for Advanced Study, Princeton,
  NJ 08540\\
    E-mail: \email{jnb@ias.edu}}
\author{Carlos Pe\~na-Garay\\
        School of Natural Sciences, Institute for Advanced Study,
  Princeton, NJ 08540\\
       E-mail: \email{penya@ias.edu}}
\abstract{We analyze all available solar and related reactor neutrino experiments, as well as simulated
future $^7$Be, $p-p$, and $pep$ solar neutrino experiments. We treat all solar neutrino fluxes as free
parameters subject to the condition that the total luminosity represented by the neutrinos equals the
observed solar luminosity (the `luminosity constraint').  Existing experiments plus the luminosity
constraint show that the $p-p$ solar neutrino flux is $1.02 \pm 0.02\, (1\sigma)$ times the flux
predicted by the BP00 standard solar model; the $^7$Be neutrino flux is
 $0.93^{+0.25}_{-0.63}$ the
predicted flux; and the $^8$B flux is $1.01 \pm 0.04$ the predicted flux. The CNO fluxes are very poorly
determined. The neutrino oscillation parameters are: $\Delta m^2 = 7.3^{+0.4}_{-0.6}\times 10^{-5}~{\rm
eV^2}$ and $\tan^2 \theta_{12} = 0.41 \pm 0.04$. We evaluate how accurate future experiments must be to
determine more precisely neutrino oscillation parameters and solar neutrino fluxes, and to elucidate the
transition from vacuum-dominated to matter-dominated oscillations at low energies. A future $^7$Be
$\nu-e$ scattering experiment accurate to $\pm 10$\% can reduce the uncertainty in the experimentally
determined $^7$Be neutrino flux by a factor of four and the uncertainty in the $p-p$ neutrino flux by a
factor of 2.5 (to $\pm 0.8$\%). A future $p-p$ experiment must be accurate to better than $\pm 3$\% to
shrink the uncertainty in $\tan^2 \theta_{12}$ by more than 15\%. The idea that the Sun shines because of
nuclear fusion reactions can be tested accurately by comparing the observed photon luminosity of the Sun
with the luminosity inferred from measurements of solar neutrino fluxes. Based upon quantitative analyses
of present and simulated future experiments, we answer the question: Why perform low-energy solar
neutrino experiments?}

\keywords{Solar and Atmospheric Neutrinos, Neutrino and Gamma
Astronomy, Beyond Standard Model, Neutrino
Physics}

\begin{document}
\input psfig

\section{Introduction}
\label{sec:introduction}

We assess in this paper how well existing and future experiments can determine neutrino parameters and
solar neutrino fluxes. The starting point for our calculations is the assumption that the large mixing
angle (LMA) MSW solution~\cite{msw} describes exactly the behavior of solar neutrino oscillations and
that the true values for the oscillation parameters lie close to the current best-estimate oscillation
parameters~\cite{postkamland,analysispostkamland} \footnote{The data included in the analysis in the
present paper include all relevant solar and reactor neutrino data available on or before September 7,
2003, the date on which the SNO collaboration announced the results of their measurements in the salt
phase and on which the GNO and SAGE collaborations announced refinements of their measurements. The
recent SNO, GNO, and GALLEX data, which are included in the analyses reported in this paper, were all
presented at the VIIIth International Conference on Topics in Astroparticle and Underground Physics
(TAUP03), Seattle (Sept. 5--9, 2003). Only minor numerical changes resulted from this updating, but we are
grateful to the referee and to JHEP for allowing us to include the new data in the final version of this
paper. An energetic reader can learn interesting and encouraging details about the robustness of current
solar neutrino inferences by comparing the many numerical results presented in hep-ph/0305159,v2. with
the numerical results presented in the present version of this paper.}.

The conclusions derived in this paper are based upon 152 global analyses of all of the available solar
neutrino and reactor data. The individual analyses differ, for example, in the input data (including
cases before and after the recent announcement of the SNO salt-phase data and the improved GNO and SAGE
measurements), the different constraints placed on the neutrino fluxes (e.g., with or without the
luminosity constraint), the number of neutrino fluxes that are treated as free variables, and the
different assumptions made about future solar neutrino and reactor experiments.   The purpose of this
introduction is to set the stage for the many detailed results that are presented in the subsequent
sections and to provide the reader with a overall map of what is contained where in this paper.

The new experiments discussed in this paper will test whether the LMA neutrino oscillation solution is
correct and, if it is correct, will determine accurately the solar neutrino fluxes, the solar neutrino
mixing angle, and the difference of the squares of the neutrino masses. We will not be surprised if these
experiments reveal physics or astronomy that cannot be explained within the now conventional framework of
the LMA oscillation solution and the standard solar model. The fundamental goal of our paper is to
determine what is expected in order to make it  easier to recognize what is unexpected.

We also show that neutrino experiments alone can be used to measure the  current rate of nuclear energy
generation in the Sun and to compare this neutrino-measured solar luminosity with the conventional
photon-measured solar luminosity. This comparison will provide an accurate test  of the fundamental
assumption that nuclear fusion among light elements is responsible for the solar luminosity. Moreover,
this same comparison will test a basic result of the standard solar model, namely, that the Sun is in a
quasi-steady state in which the current energy generation in the interior equals the current
luminosity at the solar surface.  The heat being radiated today from the solar surface  was created in
the solar interior about 40,000 years ago~\cite{diffusiontime}.

Although most of our effort in this paper is directed toward new experiments, the most surprising result
is that the $p-p$ solar neutrino flux is determined to $\pm 2$\% by the existing solar and reactor
experiments if one imposes the constraint that the luminosity of the Sun is produced by fusion reactions
among light elements that also produce solar neutrinos (the so-called `luminosity constraint'). The
best-fit value of the $p-p$ flux inferred from a global fit to existing neutrino experiments and the
luminosity constraint is within 2\% of the flux predicted by the BP00 standard solar model.

\subsection{Historical perspective}
\label{subsec:perspective}

The first forty years of solar neutrino research has demonstrated that  new physics may appear when we
carry out neutrino experiments in a new domain of sensitivity. Most of the new experiments considered in
this paper will be sensitive to neutrino energies that are less than or of the order of 1 MeV, a domain
in which solar neutrino energies could not previously be measured. More than 98\% of the predicted flux
of solar neutrinos lies below 1 MeV.

One of the primary reasons for carrying out the new experiments is to test whether new physics shows up
at lower energies. As stated earlier, our explorations in this paper are intended as a guide to the
expected in order that they can be used to help identify the unexpected.

Moreover, we want to use solar neutrino experiments for their original purpose~\cite{johnray}: to test
theoretical models of how a main sequence star gains energy and evolves by burning hydrogen to helium. To
do this, we must have sufficient experimental data to measure the total neutrino fluxes that are produced
by the principal neutrino-producing nuclear reactions. Until very recently, it was necessary to assume
the standard solar model predictions for all the solar neutrino fluxes and their uncertainties in order
to determine reasonably constrained values for neutrino oscillation parameters. Only after the results of
the Super-Kamiokande and SNO experiments became available was it possible to measure directly the $^8$B
neutrino flux. However, even now, the other solar neutrino fluxes must be taken from standard solar model
calculations in order to obtain the well-constrained global solutions for neutrino oscillation parameters
that are seen so frequently in the literature~\cite{postkamland,analysispostkamland}.

A study similar to what we carry out in this paper was performed in 1996; the earlier investigation
provides historical perspective. In 1996, only the chlorine~\cite{chlorine}, SAGE~\cite{sage02},
GALLEX~\cite{gallex} and Kamiokande~\cite{kamiokande} experimental results were available. In a paper
entitled~\cite{bk96} `How well do we (and will we) know solar neutrino fluxes and oscillation
parameters?', the authors found seven years ago that the small mixing angle (SMA) solution was slightly
preferred over the LMA and vacuum oscillation solutions. All three, SMA, LMA, and vacuum oscillations,
were allowed at $1\sigma$. The uncertainties in the solar neutrino fluxes were large.  At 95\% C.L., the
uncertainties were a factor of two for the $p-p$ neutrino flux and a factor of five for the $^8$B
neutrino flux. The $^7$Be solar neutrino flux could be as large as 6.4 times the 1995 standard solar
model prediction. In this investigation, there were only enough experiments to permit treating one flux
at a time as a free parameter. The other neutrino fluxes and their uncertainties were required to be
consistent with the 1995 standard solar model~\cite{bp95}. The principal constraints on the $p-p$ and
$^7$Be neutrino fluxes were established by the luminosity constraint~\cite{luminosity}, not the then
existing solar neutrino experiments. In 2001, Garzelli and  Giunti~\cite{giuntigarzelli} performed a
Bayesian analysis independent from the Standard Solar Model (SSM) predictions for the solar fluxes. The
luminosity constraint was used to obtain the posterior distributions of $p-p$, $^7$Be and $^8$B neutrino
fluxes in agreement with BP00 predictions~\cite{bp00}. At 90\% C.L., the allowed ranges were [0.99,1.08],
[0.02,1.15], [0.62,1.22] for $p-p$, $^7$Be, and $^8$B neutrino fluxes (in units of BP00), respectively.

With the extensive results of three additional experiments, Super-Kamiokande~\cite{superk},
SNO~\cite{snoccnc,snodaynight}, and KamLAND~\cite{kamlandfirstpaper},  we now know--or think we know--a
great deal more. But, the enormous change in the consensus view about solar neutrinos that has been
caused by the data from just three new experiments provides an additional caution that surprises may
appear in the future.

\subsection{The theme of this paper}
\label{subsec:theme}

The theme of this paper can be stated simply: the astrophysics of the solar interior and the physics of
neutrino propagation both deserve to be explored independently, not assumed. We investigate the extent to
which existing data and future experiments constrain separately both the stellar astronomy and the
neutrino physics.

A key step in our analysis is the imposition of the luminosity constraint~\cite{luminosity,spirovignaud},
which implements in a global way for the Sun the constraint of conservation of energy for nuclear fusion
among light elements. Each neutrino flux is associated with a specific amount of energy released to the
star and therefore a particular linear combination of the solar neutrino fluxes is equal to the solar
luminosity (in appropriate units). One can write the luminosity constraint as
\begin{equation}
{L_\odot\over 4\pi (A.U.)^2} = \sum\limits_i \alpha_i \Phi_i~, \label{eq:genconstraint}
\end{equation}
where $\L_\odot$ is the solar luminosity measured at the earth's surface, 1 $A.U.$ is the average
earth-sun distance,  and the coefficient $\alpha_i$ is the amount of energy provided to the star by
nuclear fusion reactions associated with each of the important solar neutrino fluxes, $\Phi_i$. The
coefficients $\alpha_i$ are calculated accurately in ref.~\cite{luminosity}.

 In this paper, we present results that are obtained with and
without imposing the luminosity constraint, in order to illustrate the power of the constraint.

 We include in our global analyses reactor~\cite{kamlandfirstpaper} and solar neutrino
data~\cite{chlorine,sage02,gallex,gno,superk,snoccnc,snodaynight}. We marginalize all the results quoted
in this paper over the mixing angle $\theta_{13}$ using the calculated dependence of the global $\chi^2$
as a function of the dominant neutrino oscillation parameters, $\Delta m^2$ and $\theta_{12}$, the solar
neutrino fluxes, and $\theta_{13}$. For earlier calculations of the dependence of  $\chi^2$ on
$\theta_{13}$, see refs. \cite{concha00,concha02,fogli02},  especially figure~7 of ref.~\cite{concha02}
and figure~4 of ref.~\cite{fogli02}. We use in this paper the dependence shown in
figure~\ref{fig:theta13} that is adapted from ref.~\cite{3nuupdate}.

 We investigate in this paper the extent to which the inferred values of the neutrino oscillation
parameters are dependent upon the common assumption that the solar neutrino fluxes have the best-estimate
values and uncertainties determined from the standard solar model~\cite{bp00}.  Throughout this paper, we
present results in which the $^8$B solar neutrino flux is treated as a free parameter. In the latter
sections of the paper, we present results in which also the $^7$Be solar neutrino flux, and later the
$p-p$ and CNO neutrino fluxes, are treated as free parameters.

We begin our investigation by determining how well the existing set of solar and reactor neutrino
experiments constrain the solar neutrino fluxes and the solar neutrino oscillation parameters. Most of
this paper is, however, devoted to experiments that are currently being planned or developed~\cite{lownu}.
We explore how well future solar neutrino experiments can be expected to determine neutrino fluxes and
oscillation parameters, all provided there are no major new surprises.

\subsection{Organization of this paper}
\label{subsec:organization}

\subsubsection{Advice for generalists  and specialists}
\label{subsubsec:advice}

A quick glance at the table of contents will be useful for both generalists and specialists. We have
written the titles of the sections and the subsections so that one can easily see what subjects are
covered and in what context.

General readers will find everything they want to learn from this paper in  our discussion section,
section~\ref{sec:discussion}. We answer in section~\ref{subsec:whylowenergy} the question: Why do low
energy solar neutrino experiments? And in section~\ref{subsec:principalresults}, we summarize the
principal results and ideas that are discussed in the main text. We give our view of what does it all
mean in section~\ref{subsec:whatmean}.

There are many detailed results and explanations that could not be included in
section~\ref{sec:discussion}. For the reader who is a specialist in a particular area related to solar
neutrino research, we nevertheless recommend  reading first the discussion section~\ref{sec:discussion}
and only afterwards selecting a particular topic from the list of section topics described below or from
the table of contents.

\subsubsection{Topics discussed in each section}
\label{subsubsec:topics}

We present in section~\ref{sec:mswvacuum} the approximate analytic behavior of the electron-neutrino
survival probability, $P_{\rm ee}$,  for the LMA oscillation solution. A number of the results in this
paper can be understood qualitatively by referring to figure~\ref{fig:survival} of
section~\ref{sec:mswvacuum}.

Section~\ref{sec:technical} describes some of the technical details that we use later in the paper,
including the input experimental data, the ingredients of the global $\chi^2$, the relationship between
the two-neutrino and three neutrino survival probabilities, and the simulations we make of future KamLAND
and SNO data. The details of section~\ref{sec:technical} are primarily of interest to experts on neutrino
oscillations and can be skipped by the general reader who just wants to know the scientific implications
of our investigations.

We use in section~\ref{sec:SplusK} all of the available solar and reactor neutrino data to determine the
current constraints on neutrino oscillation parameters and solar neutrino fluxes (see especially
table~\ref{tab:rangeSplusKppb8be7} and figure~\ref{fig:SplusK}). The global analyses in this section
evolve in steps from treating only the $^8$B neutrino flux as free, to letting both the $^7$Be and $^8$B
fluxes be free parameters, to the most powerful method: treating all of the neutrino fluxes as free
parameters subject to the luminosity constraint.  We also analyze in this section simulated data for
three years of operation of KamLAND and show how these additional data are expected to improve our
knowledge of oscillations parameters and fluxes.

We explain in section~\ref{sec:luminosityexplain} the reason why the luminosity constraint helps provide
a powerful bound on the $p-p$ solar neutrino flux,  but imposes only a relatively weak constraint on the
$^7$Be solar neutrino flux.

Section~\ref{sec:whatbe7} and section~\ref{sec:whatpp} provide our best answers to the questions that
experimentalists planning future solar neutrino measurements most often ask us.  These sections address
the parallel questions: What will we learn from a $^7$Be solar neutrino experiment? What will we learn
from a $p-p$ solar neutrino experiment?  In order to answer these questions, we analyze simulated data
for $^7$Be and $p-p$ solar neutrino experiments.

We show in section~\ref{sec:whatbe7}  that the current predictions for a $^7$Be solar neutrino experiment
depend sensitively on whether or not one assumes the correctness of the best-estimate neutrino fluxes,
and their uncertainties, that are predicted by the standard solar model. Table~\ref{tab:rangeslmanycases}
shows that we will be stuck with a large experimental uncertainty in the flux of $^7$Be solar neutrinos
unless a $^7$Be solar neutrino experiment accurate to $\sim \pm 5$\% is performed.  Fortunately, a $^7$Be
solar neutrino experiment will also provide a powerful constrain on the $p-p$ solar solar flux via the
luminosity constraint.

We show in section~\ref{sec:whatpp}  that a measurement of the $p-p$ neutrino-electron scattering rate
must be very accurate, $\sim \pm 1$\%, in order to significantly improve our knowledge of the $\tan ^2
\theta$ or the solar neutrino fluxes (see especially table~\ref{tab:freepp}). If this accuracy is
achieved, then the allowed range for $\tan^2 \theta_{12}$ will be reduced by a factor of two. We
demonstrate also in this section that a measurement of the $pep$ neutrino flux would be about as
informative as a measurement of the $p-p$ neutrino flux.

We answer in section~\ref{sec:maximalmixing} two important questions: At what level is maximal mixing
excluded by current data?  How will the SNO salt-phase data affect the exclusion of maximal mixing?

\section{MSW at high energies, Vacuum at low energies}
\label{sec:mswvacuum}

The effective Hamiltonian for two-neutrino propagation in matter can be written conveniently in the
familiar form~\cite{msw,bethe,Mikheev:ik,messiah,neutrinoastrophysics,conchayossi}

\begin{equation}
H ~=~ \left ( \begin{array}{cc} \frac{\Delta m^2 cos 2 \theta_{12}}{4 E}- \frac{\sqrt{2}G_{\rm F} n_{\rm
e}}{2}&
\frac{\Delta m^2 sin2 \theta_{12}}{2 E}\\
 \frac{\Delta m^2 sin2 \theta_{12}}{2 E} &
 -\frac{\Delta m^2 cos 2 \theta_{12}}{4 E}+ \frac{\sqrt{2}G_{\rm F} n_{\rm e}}{2}\end{array}\right) \, .
 \label{eq:hamiltonian}
\end{equation}
Here $\Delta m^2$ and $\theta_{12}$ are, respectively, the difference in the squares of the masses of the
two neutrinos and the vacuum mixing angle, $E$ is the energy of the neutrino, $G_{\rm F}$ is the Fermi
coupling constant, and $n_{\rm e}$ is the electron number density at the position at which the
propagating neutrino was produced. The best-fit values for $\Delta m^2$ and $\tan^2 \theta_{12}$ obtained
from a global solution to all currently available solar neutrino and reactor anti-neutrino experimental
data are (see table~\ref{tab:rangeSplusKppb8be7}, the third row) $\Delta m^2 = 7.3 \times 10^{-5} \, {\rm
eV^2}$ and $\tan^2 \theta_{12} = 0.41$.

The  relative importance of the MSW matter term and the kinematic vacuum oscillation term in the
Hamiltonian can be parameterized by the quantity, $\beta$, which represents the ratio of matter to vacuum
effects. From equation~\ref{eq:hamiltonian} we see that the appropriate ratio is
\begin{equation}
\beta= \frac{2 \sqrt2 G_F n_e E_\nu}{\Delta m^2}\, . \label{eq:defbeta}
\end{equation}
The quantity $\beta$ is the ratio between the oscillation length in matter and the oscillation
 length in vacuum. In convenient units,
$\beta$ can be written as
\begin{equation}
\beta= 0.22 \, \left[\frac{E_\nu}{1~{\rm MeV}}\right]\, \left[ \frac{\mu_e\rho}{100~{\rm
g~cm}^{-3}}\right] \, \left[ \frac{7 \times 10^{-5} eV^2}{\Delta m^2}\right]\, ,
\label{eq:betaconvenient}
\end{equation}
where $\mu_e$ is the electron mean molecular weight ($\mu_e \approx 0.5(1 + X)$, where X is the mass
fraction of hydrogen) and $\rho$ is the total density, both evaluated at the location where the neutrino
is produced. For the electron density at the center of the standard solar model, $\beta = 0.22$ for $ E =
1$MeV and $\Delta m^2 =  7\times 10^{-5} {\rm eV^2}$.

 For the LMA region, the daytime survival probability can be written to a  good approximation in the
following simple form~\cite{msw,bethe,Mikheev:ik,messiah,neutrinoastrophysics,parke}
\begin{equation}
P_{\rm ee}~=~  \frac{1}{2} ~+~  \frac{1}{2} \cos2\theta^M_{12} \cos2\theta_{12}\, , \label{eq:plmaday}
\end{equation}
where the mixing angle in matter is
\begin{equation}
\cos2\theta^M_{12} = \frac{\cos2\theta_{12}-\beta}{\sqrt{(\cos2\theta_{12}-\beta)^2+\sin^22\theta_{12}}}\, .
\label{eq:defthetaM}
\end{equation}

Figure~\ref{fig:survival} illustrates the energy dependence of the LMA survival probability, $P_{\rm
ee}$. If $\beta < \cos 2\theta_{12} \sim 0.4$ (for solar neutrino oscillations), the  survival probability
corresponds to vacuum averaged oscillations,
\begin{equation}
P_{\rm ee}~=~ 1 - \frac{1}{2} \sin^22\theta_{12} ~\, (\beta < \cos 2\theta_{12}, ~{\rm vacuum})
.\label{eq:peevacuum}
\end{equation}
If $\beta > 1$, the  survival probability corresponds to matter dominated oscillations,
\begin{equation}
P_{\rm ee}~=~ \sin^2\theta_{12}~\, (\beta > 1, ~{\rm MSW}).
\end{equation}
The survival probability is approximately constant in either of the two limiting regimes, $\beta < \cos
2\theta_{12}$ and $\beta > 1$. The LMA solution exhibits strong energy dependence only in the transition
region between the limiting regimes.

\FIGURE[!ht]{ \centerline{\psfig{figure=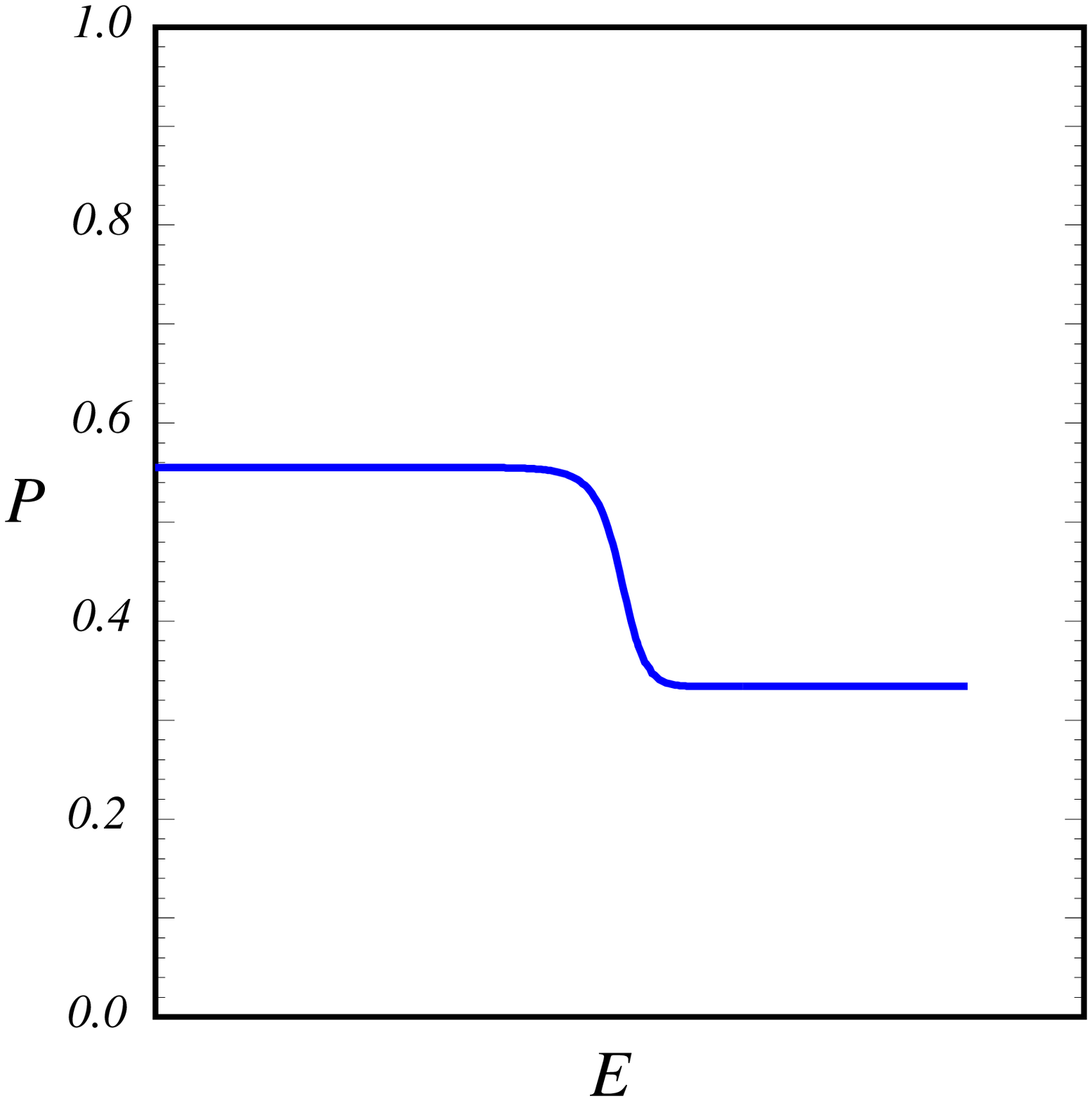,width=3.5in}} \caption{{\bf Schematic survival
probability.} The figure shows the electron neutrino survival probability, $P_{\rm ee}$, as a function of
neutrino energy for the (daytime) LMA oscillation solution.  For small values of the parameter $\beta$
defined in equation~\ref{eq:defbeta} and equation~\ref{eq:betaconvenient}, the kinematic (vacuum)
oscillation effects are dominant. For values of $\beta$ greater than unity, the MSW (matter) oscillations
are most important. For solar conditions, the transition between vacuum and matter oscillations occurs
somewhere in the region of 2 MeV. \label{fig:survival}}}

At what neutrino energy does the transition take place between vacuum oscillations and matter
oscillations? The answer to this question depends upon which neutrino source one discusses, since the
fraction of the neutrino flux that is produced at a given radius (i.e., density and $\mu_e$) differs from
one neutrino source to another. The $^8$B neutrinos are produced at much smaller radii (higher densities)
than the $p-p$ neutrinos; the $^7$Be production profile is intermediate between the $^8$Be and $p-p$
neutrinos.  According to the BP00 solar model, the critical energy at which $\beta = \cos 2\theta_{12}$
is, for $\tan^2 \theta_{12} = 0.41$,
\begin{equation}
E({\rm crit})~\simeq 1.8 {\rm ~ MeV\,(^8B)}; ~~\simeq 2.2 {\rm ~ MeV \,(^7Be)}; ~~\simeq 3.3 {\rm ~ MeV
\,}(p-p).
 \label{eq:critical}
\end{equation}

The actual energies for $p-p$ and $^7$Be neutrinos are below the critical energy where they are produced.
To a very good approximation, $^8$B neutrinos are always in the MSW regime, while $p-p$ and $^7$Be
neutrinos are in the vacuum regime. The energies are so low for $p-p$ and $^7$Be neutrinos,
that the energy dependence of the vacuum oscillations, $\propto \sin^2(\Delta m^2 L/4E)$, averages very close
to one-half in any measurable energy interval. To a good approximation, the survival probability is given by
equation~\ref{eq:peevacuum} for both $^7$Be and $p-p$ neutrinos.

The fact that the survival probability in equation~\ref{eq:peevacuum}  is independent of $\Delta m^2$
explains why the $^7$Be and $p-p$ experiments discussed in section~\ref{sec:whatbe7} and
section~\ref{sec:whatpp} are insensitive to $\Delta m^2$. We shall show in section~\ref{sec:whatpp} that
a precise $p-p$ experiment can be sensitive to $\tan^2 \theta_{12}$.

 The rather low critical energy for $^8$B neutrinos
explains why spectral distortions are difficult to measure in Super-Kamiokande and SNO with present
thresholds $E_e \sim {5 \rm ~MeV}$. However, a distortion should be measurable if the threshold can be
lowered by 1-2 MeV.

\section{Some technical remarks}
\label{sec:technical}

In this section, we describe some technical aspects of our analysis and introduce some of the notation.
This section can be skipped by  readers who do not suffer from  acute addiction to the details of
neutrino oscillation analyses. Where necessary in the remainder of the text, we refer back to this
section for definitions or for specific procedures.

We summarize in section~\ref{subsec:experimental} (see especially table~\ref{tab:experimental}) the
existing experimental data used in latter sections to carry out global analyses of solar and reactor
neutrino experiments. In section~\ref{subsec:globalchi}, we describe the global $\chi^2$ we have used and
define the free parameters in $\chi^2_{\rm global}$. In particular, we define in
section~\ref{subsec:globalchi} the reduced neutrino fluxes, $f_{\rm B}$ ($^8$B neutrinos), $f_{\rm Be}$
($^7$Be neutrinos), and $f_{p-p}$ ($p-p$ neutrinos). We define in section~\ref{subsec:survival} the
effective survival probability after marginalizing over $\theta_{13}$. In
section~\ref{subsec:simulatedkamland}, we describe how we simulate KamLAND data for three years of
operation. Finally, in section~\ref{subsec:simulationsno} we describe how we simulate the SNO data that
will result from the salt-phase measurements.

\subsection{Experimental data}
\label{subsec:experimental}

Table~\ref{tab:experimental} summarizes the
solar (S)~\cite{chlorine,sage02,gallex,snoccnc,snodaynight,gno} and KamLAND (K)
data~\cite{kamlandfirstpaper} used in our global analyses that are presented latter in this paper. The
total number of data used in the analyses is 96; the number of data derived from each
experiment are listed (in parentheses) in the second column of the
table.  In the third column, labeled Measured/SM, we list for each experiment the quantity Measured/SM, the
measured total rate divided by the rate that is expected based upon the standard solar model and the standard
model of electroweak interactions.

\begin{table}
\small \caption[]{{\bf Experimental data.} We summarize the solar (S) and
the KamLAND (K)
data~\cite{kamlandfirstpaper} used in our global analyses. Only experimental errors are included in the
column labeled Result/SM. Here the notation {SM} corresponds to predictions of the Bahcall-Pinsonneault
standard solar model (BP00) of ref.~\protect\cite{bp00}  and the standard model of electroweak
interactions~\cite{sm} (with no neutrino oscillations). The SNO rates (pure $D_2O$ phase) in the column
labeled
Result/SM are
obtained from the published SNO spectral data by assuming that the shape of the ${\rm ^8B}$ neutrino
spectrum is not affected by physics beyond the standard electroweak model. However, in our global
analyses, we allow for spectral distortion. The SNO rates (salt phase) are not constrained to the $^8$B
shape \protect\cite{snosalt}.
\label{tab:experimental}
\\}
\begin{tabular}{lcccc}
\hline\noalign{\smallskip}
Experiment & Observable ($\#$ Data) & Measured/SM & Reference\\
\noalign{\smallskip}\hline\noalign{\smallskip}
 Chlorine & Average Rate (1) & [CC]=$0.34 \pm 0.03$ &\ \ \protect\cite{chlorine}\\
%\hline
SAGE+GALLEX/GNO$^{\dagger}$ & Average Rate (1)& [CC]=$0.54 \pm 0.03$ &\ \ \protect\cite{sage02,gallex,gno}\\
%\hline
Super-Kamiokande & Zenith Spectrum (44)& [ES]=$0.465 \pm 0.015$ &\ \ \protect\cite{superk}\\
%\hline
SNO (pure D2O phase) & Day-night Spectrum (34)&[CC]=$0.35 \pm 0.02$&\ \ \protect\cite{snoccnc,snodaynight}\\
 & &[ES]=$0.47 \pm 0.05$&\ \ \cite{snoccnc,snodaynight}\\
 & &[NC]=$1.01 \pm 0.13$&\ \ \cite{snoccnc,snodaynight}\\
%\hline
SNO (salt phase) & Average Rates (3)&[CC]=$0.32 \pm 0.02$&\ \ \protect\cite{snosalt}\\
 & &[ES]=$0.44 \pm 0.06$&\ \ \cite{snosalt}\\
 & &[NC]=$1.03 \pm 0.09$&\ \ \cite{snosalt}\\
%\hline
KamLAND & Spectrum (13)&[CC]=$0.61 \pm 0.09$&\ \ \protect\cite{kamlandfirstpaper}\\
\noalign{\smallskip}\hline\noalign{\smallskip}
\end{tabular}
\hbox to\hsize{$^{\dagger}$ SAGE rate: $69.1^{+4.3}_{-4.2}~^{+3.8}_{-3.4}$~SNU~\protect\cite{sage02};
GALLEX/GNO rate: $69.3 \pm 4.1 \pm 3.6$~SNU~\protect\cite{gallex,gno}. \hfill}

\end{table}

\subsection{Global $\chi^2$, free parameters, and reduced solar neutrino fluxes}
\label{subsec:globalchi}

We calculate the global $\chi^2$ by fitting to all the available data, solar
(83 measurements) plus
reactor (13 measurements). Formally, the global $\chi^2$ can be written in the form~\cite{chi2a,chi2b}

\begin{eqnarray}
\chi^2_{\rm global} &=& \chi^2_{\rm solar}(\Delta m^2,\theta_{12}, \{\xi, f_{\rm B}, f_{\rm Be},
f_{p-p}, f_{\rm CNO}\}) \nonumber\\
 &+& \chi^2_{\rm KamLAND}(\Delta m^2,\theta_{12},\theta_{13})
~+~ \chi^2_{\rm CHOOZ+ATM}(\theta_{13}) \,.
\label{eq:chisquaredpowerful}
\end{eqnarray}

Depending upon the case we consider, there can be as many as nine free parameters in $\chi^2_{\rm
solar}$, including, $\Delta m^2, \theta_{12}$, $\xi$ , $f_{\rm B},$ $f_{\rm Be}$, $f_{p-p}$, and $f_{\rm
CNO}$ (3 CNO fluxes, see below). The neutrino oscillation parameters $\Delta m^2, \theta_{12}$ have their
usual meaning. The reduced fluxes $f_{\rm B}$, $f_{\rm Be}$, $f_{p-p}$, and $f_{\rm CNO}$ are defined
below. The parameter $\xi$ describes departures from two-neutrino oscillations. In general, $\xi$ is a
vector defined by several parameters. If we consider the standard scenario of three flavors, the higher
mass squared averages out and $\xi\equiv\theta_{13}$.

The function $\chi^2_{\rm KamLAND}$ depends only on $\Delta m^2$, $\theta_{12}$ and $\theta_{13}$. We
discuss in section 2.2 of ref.~\cite{postkamland} our procedure for analyzing the data from the KamLAND
experiment.

For most of the calculations described in this paper, we marginalize $\chi^2_{\rm global}$ making use of
the function $\chi^2_{\rm CHOOZ+ATM}(\theta_{13})$  that was obtained in the updated
analysis~\cite{3nuupdate} of atmospheric~\cite{atm}, K2K accelerator~\cite{k2k}, and CHOOZ
reactor~\cite{chooz,paloverde}  data (see also, refs.~\cite{concha02,concha00,fogli02}).
Figure~\ref{fig:theta13} (which is the same as Fig.~3c of ref.~\cite{3nuupdate}) shows the specific
dependence upon $\theta_{13}$ that we use.  This formulation takes account of all the available data from
solar, reactor, and accelerator neutrino oscillation experiments. We have not assumed, as is often done,
a flat probability distribution for all values of $\theta_{13}$ below the CHOOZ bound. The fact that we
take account of the actual experimental constraints on $\theta_{13}$ decreases the estimated influence of
$\theta_{13}$ compared to what would have been obtained for a flat probability distribution.

\FIGURE[!t]{ \centerline{\psfig{figure=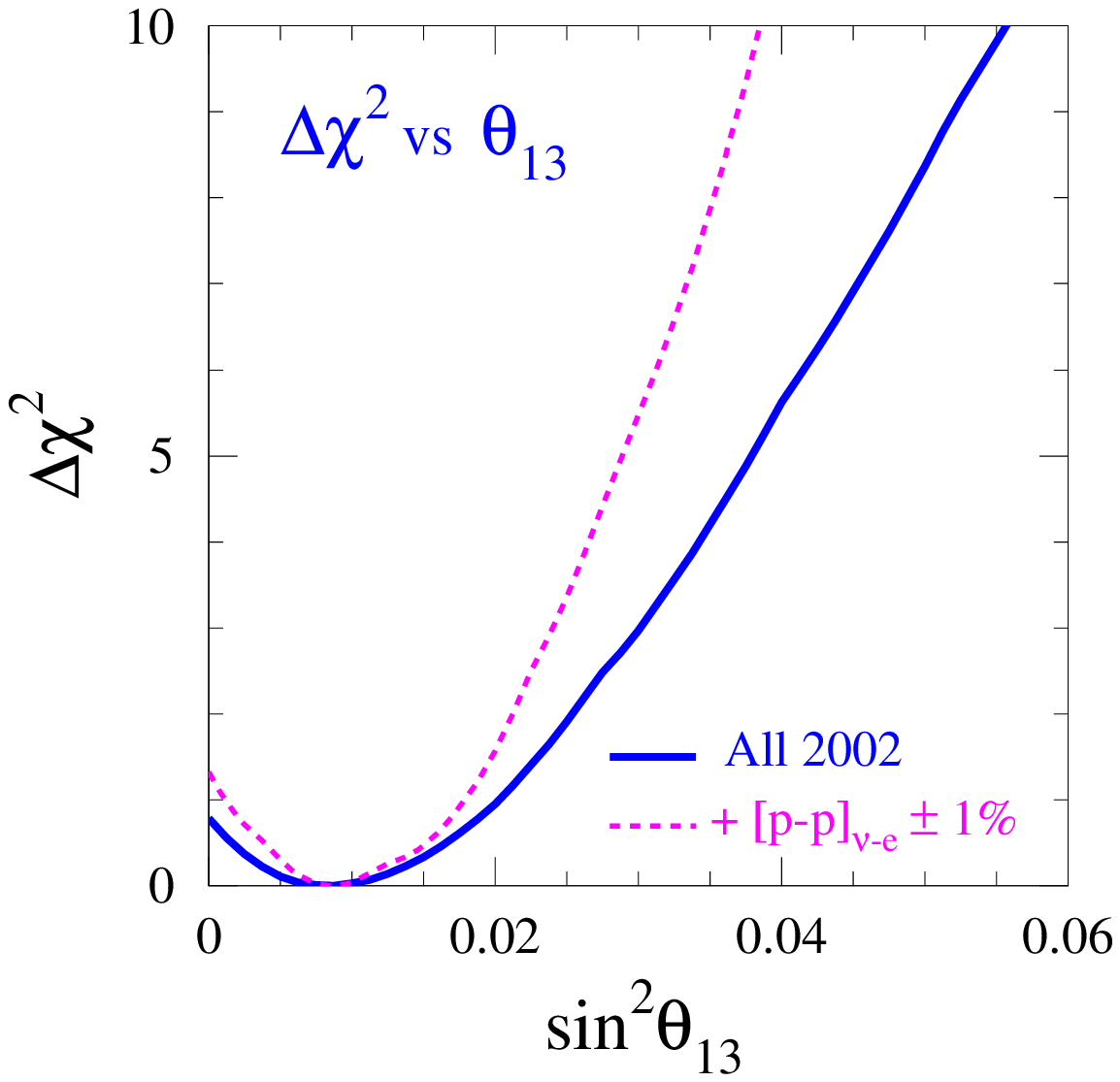,width=3.5in}} \caption{{\bf $\Delta \chi^2$ dependence
on $\theta_{13}$.} Here $\theta_{13}$ is the mixing angle between the first and the third neutrino
eigenstates. The dependence of $\chi^2$ upon $\theta_{13}$  that is shown in the figure corresponds to
the global analysis of solar+atmospheric+KamLAND+K2K+CHOOZ data available in 2002 and displayed in
Fig.~3c of ref.~\cite{3nuupdate}. We have used the 2002 dependence shown in the figure in calculating the
global $\chi^2$ defined by equation~\ref{eq:chisquaredpowerful}. We have not assumed a flat distribution
for $\theta_{13}$ below the CHOOZ bound. If a 1\% $p-p$ solar neutrino experiment is performed and
analyzed in conjunction with a 5\% $^7$Be solar neutrino experiment, three years of KamLAND reactor data,
and  the currently available solar neutrino data, then the bound in the curve labeled ` + $p-p$ ' can be
achieved.  \label{fig:theta13}}}

We have included in our analysis the possibility of an active-sterile admixture and in that case
$\xi\equiv\eta$, where $\eta$ describes the possibility that $\nu_e$ oscillates into a state that is a
linear combination of active ($\nu_a$) and sterile ($\nu_s$) neutrino states,
\begin{equation}
\nu_e \to \cos\eta \, \nu_a \,+\, \sin\eta \,\nu_s \, . \label{eq:linearcombination}
\end{equation}
 This admixture arises in the framework of 4-$\nu$ mixing \cite{four}. The analysis procedure we follow
 in the case of a non-zero admixture of sterile neutrinos is described in section~3.3.2 of
 ref.~\cite{postkamland}. For this case only, we assume $\theta_{13} = 0$.

We follow the usual convention and represent the reduced $^8$B solar neutrino flux by the parameter
$f_{\rm B}$ here $f_{\rm B}$ is the ratio of the true flux to the flux predicted by the standard solar
model. Thus
\begin{equation}
f_{\rm B} ~\equiv~ \frac{\phi(^8{\rm B})}{~\phi(^8{\rm B})_{\rm BP00}}\,. \label{eq:fbtotaldefinition}
\end{equation}
Here $\phi(^8{\rm B})_{\rm BP00}~\equiv~5.05\times 10^6 \,{\rm cm^{-2}s^{-1}}$~\cite{bp00}. We define the
reduced $^7$Be neutrino flux, $f_{\rm Be}$, by analogy with $f_{\rm B}$. Thus
\begin{equation}
f_{\rm Be} ~\equiv~ \frac{\phi(^7{\rm Be})}{~\phi(^7{\rm Be})_{\rm BP00}}\,, \label{eq:fbedefinition}
\end{equation}
where $\phi(^7{\rm Be})_{\rm BP00}~\equiv~4.77\times 10^9\,{\rm cm^{-2}s^{-1}}$ with a $1\sigma$
uncertainty of $\pm 10$\%. The reduced $p-p$ neutrino flux is defined by
\begin{equation}
f_{p-p} ~\equiv~ \frac{\phi({p-p})}{~\phi({p-p})_{\rm BP00}}\,,
\label{eq:fppdefinition}
\end{equation}
where ~\cite{bp00} $\phi({p-p})_{\rm BP00}~\equiv~5.95\times 10^{10}\,{\rm cm^{-2}s^{-1}}$ with a
$1\sigma$ uncertainty of $\pm 1$\%.

Similarly, we define reduced $^{13}$N,$^{15}$O and $^{17}$F neutrino fluxes that we parameterize by
$f_{\rm CNO}$. We have checked that different limiting cases in the relative weights of the $^{13}$N,
$^{15}$O and $^{17}$F fluxes (see discussion in ref.~\cite{cnopaper}) do not affect the accuracy of our
results.

\subsection{Survival probability and $\theta_{13}$}
\label{subsec:survival}

 The survival probability of electron neutrinos in the case of three neutrino
oscillations, $P^{3\nu}_{ee}$, can be simply related to the survival probability, $P_{\rm ee}^{2\nu}$,
for two neutrino oscillations by the equation~\cite{3nu}
\begin{equation}
P^{3\nu}_{ee}~=~\cos^4\theta_{13} P^{2\nu}_{ee}(\Delta m^2, \theta_{12} ; n_e \to \cos^2\theta_{13} n_e)
+ \sin^4\theta_{13}.
 \label{eq:three}
\end{equation}
The effect of $\Delta M^2$, the mass difference squared characteristic of atmospheric neutrinos, averages
out  in equation~\ref{eq:three} for the energies and distances characteristic of solar neutrino
propagation. The  effective two-neutrino problem is solved with a re-normalized electron density
$\cos^2\theta_{13} n_e$. The results from the CHOOZ reactor experiment~\cite{chooz,paloverde} place a strong
upper bound on $\sin^2 2\theta_{13}$, implying that $\theta_{13}$ is close to $0$ or close to $\pi/2$.
Atmospheric and solar data select the first option ($\cos^4 \theta_{13}$ close to $1$ and $\sin^4
\theta_{13}$ close to $0$). Thus the main effect of a small allowed $\theta_{13}$ on the survival
probability is the introduction of the factor $\cos^4\theta_{13}$ in equation~\ref{eq:three}.

In what follows, we shall use `survival probability' and  $P_{ee}$ to denote $P^{3\nu}_{ee}$.  Where
numerical results are reported, we marginalize over $\theta_{13}$.  At all points in oscillation
parameter space, we use the value of $\theta_{13}$ that minimizes $\chi^2$ for that set of parameters.

\subsection{Marginalization for predicted event rates}
\label{subsec:marginalization} In order to calculate properly the allowed ranges of the reduced event
rates for  $[^7{\rm Be}]_{\nu-e}$ (defined in equation~\ref{eq:be7reducedrate}) or $[{p-p}]_{\nu-e}$
(defined in equation~\ref{eq:ppreducedrate}), we first evaluate a global minimum $\chi^2$. We then
calculate, for example, $\chi^2$ and $[^7{\rm Be}]_{\nu-e}$ at each of many points in the
multi-dimensional input parameter space. From this data set, we construct the minimum value of $\chi^2$
at each value of $[^7{\rm Be}]_{\nu-e}$.   Thus we are able to construct a function
$\chi^2_{\min}([^7{\rm Be}]_{\nu-e})$. We determine the allowed range of  $[^7{\rm Be}]_{\nu-e}$ by
selecting the values of $[^7{\rm Be}]_{\nu-e}$ that produce the appropriate $\Delta \chi^2$ for the
specified confidence limit and 1 d.o.f.

\subsection{Simulated KamLAND data}
\label{subsec:simulatedkamland}

We simulate three years of KamLAND measurements using the data provided in the KamLAND
paper~\cite{kamlandfirstpaper}. We multiply by nine the number of events for all the prompt energy bins
above 2.6 MeV.  Following the KamLAND collaboration, we limit our  analysis to energy bins above 2.6 MeV
in order to eliminate the  geo-neutrino background. The energy bins above 6 MeV contain zero events in
the first KamLAND results. For these bins, we simulate the expected signal in 3 years of KamLAND data by
using the number of events predicted by the present best fit oscillation solution ($\tan^2 \theta_{12} =
0.42$, $\Delta m^2 = 7.1 \times 10^{-5} eV^2$) for all the available solar plus KamLAND data.

\subsection{SNO salt phase data} \label{subsec:simulationsno}

In the analyses presented in this paper, we have included the salt-phase SNO data~\cite{snosalt} and the
refined GNO~\cite{gno} and SAGE~\cite{sage02} data, all released on September 7, 2003. Every one of the
analyses performed in this paper was carried out with (in the current version of the paper) and without
(in the previous version, hep-ph/0305159,v2) the new SNO, GNO, and SAGE data. Performing the analyses
with and without the new data allowed us to double check all our conclusions and to verify that the
departure from maximal mixing was the only physical variable whose value or statistical significance was
much affected by the addition of the new data\footnote{In the version of this paper that was posted on
the archive on June 19, 2003 and that was submitted for publication prior to the availability of the
recently released SNO salt-phase data, we made best-estimate simulations of the expected salt-phase
measurements (see hep-ph/0305159,v2, section 3.6). We shall see in section~\ref{subsec:snohowwell} that
these simulations turned out to be accurate and the conclusions that were deduced from the simulations
are in agreement with those derived by the SNO collaboration from the actual measurements. The agreement
between the simulations and the measurements for the SNO salt-phase provides circumstantial evidence that
the simulations we make later in this paper for other experiments actually capture the essential features
of those new experiments. Readers interested in simulating solar neutrino measurements may profit by
reading section~3.6 of hep-ph/0305159,v2 to see what are the essential ingredients of an accurate
simulation of the SNO data.}.

 In this version of our paper, we have included the charged current, neutral current, and elastic scattering rates
\cite{snosalt} as determined by the SNO collaboration, where the spectral distributions of electron
scattering and charged current events are not constrained by the known shape of the $^8$B neutrino energy
spectrum. We treated the errors in the salt phase measurements as described in the detailed user guide
written by the collaboration~\cite{howto2}; we no longer need to use our simulations of the SNO
salt-phase data.

\section{Solar plus KamLAND constraints}
\label{sec:SplusK}

What can we say today about neutrino oscillation parameters and solar neutrino fluxes using a global
analysis of all the existing data on solar and reactor neutrino experiments?  And, what improvements can
we expect in our knowledge of the oscillation parameters and neutrino fluxes after three years of data
taking by the KamLAND collaboration?

Using a global analysis, we determine in section~\ref{subsec:SplusKb8} the allowed regions of neutrino
oscillation parameters and the $^8$B solar neutrino flux assuming that the $^8$B  neutrino flux is a free
parameter. In this subsection, we assume that all solar neutrino fluxes except the $^8$B neutrino flux
are described well by the standard solar model calculation~\cite{bp00} of their best-estimates and
uncertainties. Allowing both the $^7$Be and the $^8$B neutrino fluxes to be free parameters, we repeat
the global analysis in section~\ref{subsec:SplusKbe7b8}. We discuss in
section~\ref{subsec:be7implications} the changes in the predictions for future solar neutrino experiments
that are caused by allowing the $^7$Be neutrino flux to vary freely.

\FIGURE[!t]{ \centerline{\psfig{figure=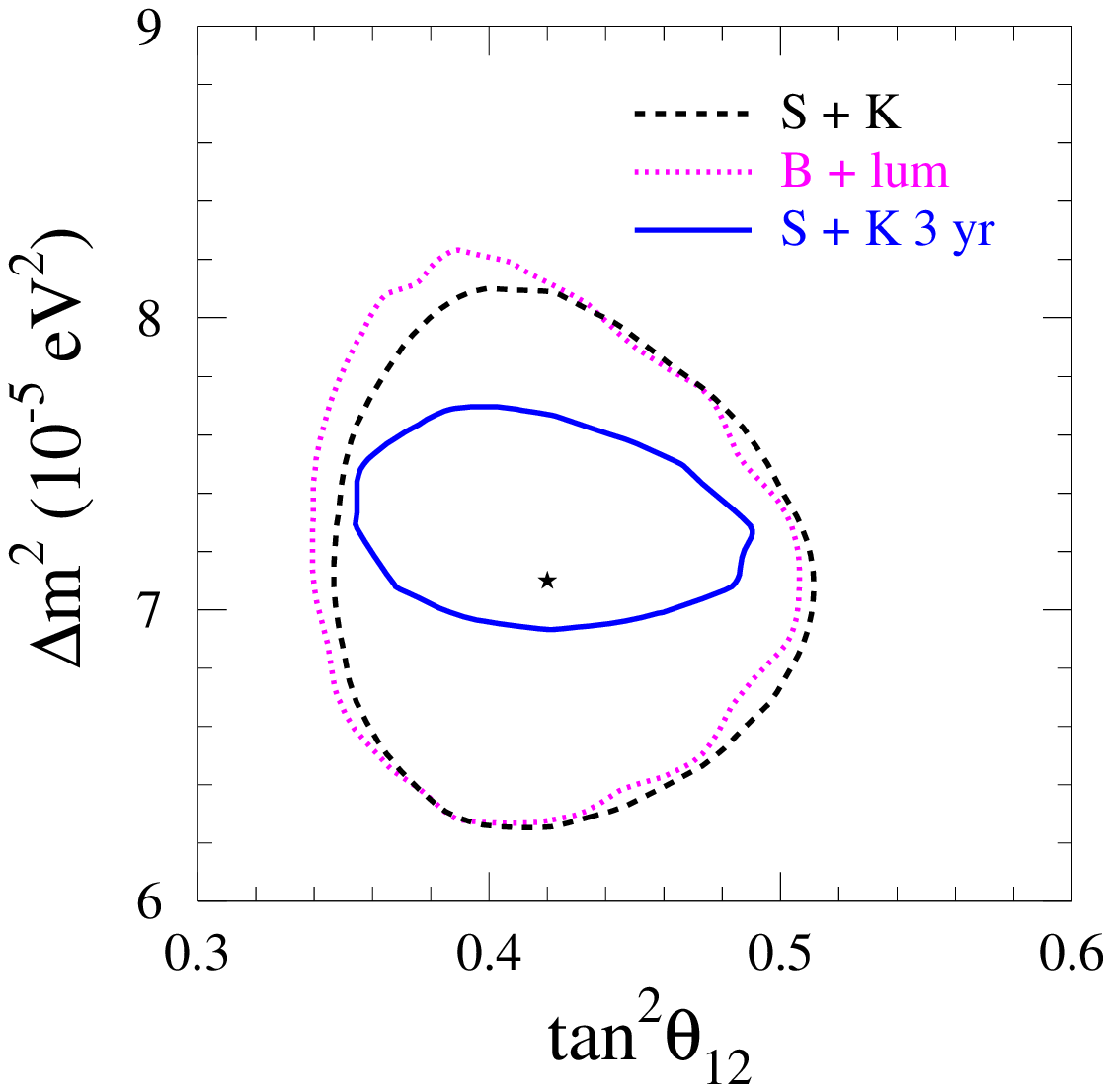,width=3.5in}} \caption{{\bf Allowed oscillation
parameters: Solar plus KamLAND measurements.} The figure shows the currently allowed $1\sigma$ region for
oscillation parameters that is obtained by a global fit, as described in the text, of all the available
solar (S)~\cite{chlorine,sage02,gallex,superk,snoccnc,snodaynight,gno,snosalt} plus reactor (K) data. At
$3\sigma$, the allowed regions form two connected islands~\cite{postkamland}. The current best-fit point
is labeled by a solid star. The figure also shows how the allowed contour is expected to be reduced after
three years of operation of the KamLAND reactor experiment.  In constructing these two contours, the
$^8$B solar neutrino flux was treated as a free parameter and all the other solar neutrino fluxes, and
their theoretical uncertainties, were taken from the standard solar model of BP00~\cite{bp00} The figure
also shows a contour labeled `B + lum', which corresponds to the case summarized in
table~\ref{tab:rangeSplusKppb8be7}, in which all the neutrino fluxes are treated as free parameters
subject to the luminosity constraint. The $1\sigma$ contours were drawn  in all cases by marginalizing
over $\theta_{13}$ and the $^8$B solar neutrino flux.  \label{fig:SplusK}}}

The most dramatic result is obtained in section~\ref{subsec:SplusKppbe7b8}.  In this subsection,  we
impose the luminosity constraint while allowing the $p-p$, $^7$Be, $^8$B, and CNO neutrino fluxes to vary
freely. The existing solar neutrino and KamLAND experiments constrain the total $p-p$ flux to $\pm 2$\%
($1\sigma$)and are in agreement with the standard solar model prediction~\cite{bp00} within the
calculated error.

We present in section~\ref{subsec:sterile} the constraints on the sterile neutrino fraction of the $^8$Be
neutrino flux that are implied by the existing solar and KamLAND data.  We also evaluate the effect of
three years of simulated KamLAND data on the expected bound for the sterile fraction of the $^8$B
neutrino flux.

We investigate in section~\ref{subsec:cno} the extent to which solar neutrino and reactor data can
constrain the CNO contribution to the solar luminosity and in section~\ref{subsec:totalluminosity} we
show that neutrino observations alone can provide a 20\% measurement of the total solar luminosity.

All our of investigations in this section show that we will be stuck with a large uncertainty in the
experimentally-determined $^7$Be solar neutrino flux until a dedicated $^7$Be solar neutrino experiment
is performed.

For the reader who is mainly interested in the bottom line, we recommend jumping directly to
section~\ref{subsec:SplusKppbe7b8}.

\subsection{$^8$B neutrino flux as a free parameter}
\label{subsec:SplusKb8}

Figure~\ref{fig:SplusK} shows the $1\sigma$ allowed region for the neutrino oscillation parameters
$\Delta m^2$ and $\tan^2\theta_{12}$ that is  obtained by a global fit to all of the existing solar plus
KamLAND data.  In addition, the figure shows the reduction in the allowed region that may be expected if
the KamLAND experiment takes data for a total of three years.

\TABLE[!t]{\caption{{\bf Allowed neutrino parameters and $^8$B solar
 neutrino flux: Solar plus KamLAND measurements.}
 For $\Delta m^2$, $\tan^2\theta_{12}$, and the $^8$B solar neutrino flux, the table presents the
 global best fit and $1\sigma$
($3\sigma$) allowed ranges. The first row shows the allowed ranges determined from a global analysis of
all existing solar neutrino (S')data available before September 7, 2003 as well as the initial  KamLAND
(K) data. The the second and third rows are determined from a global analysis of all existing solar (S)
and KamLAND (K) data, including the SNO, GNO, and SAGE data~\cite{sage02,gno,snosalt} released on
September 7, 2003. The $^8$B solar neutrino flux is treated as a free parameter, but the $^7$Be and all
other solar neutrino fluxes are assumed to have the standard solar model~\cite{bp00} predicted values and
uncertainties.\label{tab:rangeSplusK}}
\begin{tabular}{lccc}
\hline \noalign{\smallskip} Analysis & $\Delta m^2 (10^{-5} eV^2)$&$\tan^2\theta_{12}$&$f_B$\\
\hline \noalign{\smallskip}\noalign{\smallskip} S' + K (before) & $7.1^{+0.6}_{-0.4}$ ($^{+11.8}_{-1.6}$)
& $0.45^{+0.07}_{-0.06}$ ($^{+0.37}_{-0.16}$)
& $1.00^{+0.05}_{-0.06}$ ($^{+0.16}_{-0.19}$)\\
\noalign{\smallskip}\noalign{\smallskip} S  + K  (after)& $7.1^{+0.4}_{-0.4}$ ($^{+2.6}_{-1.6}$) &
$0.42^{+0.05}_{-0.04}$ ($^{+0.21}_{-0.12}$)
& $1.01^{+0.04}_{-0.04}$ ($^{+0.13}_{-0.13}$)\\
\noalign{\smallskip}\noalign{\smallskip} S + K 3 yr & $7.3^{+0.2}_{-0.2}$ ($^{+0.9}_{-0.6}$) &
$0.42^{+0.04}_{-0.04}$ ($^{+0.15}_{-0.10}$)
& $1.01^{+0.03}_{-0.03}$ ($^{+0.10}_{-0.10}$)\\
\noalign{\smallskip} \hline
\end{tabular}
}

Table~\ref{tab:rangeSplusK} shows the allowed ranges at $1\sigma$ ($3\sigma$) of the $^8$B solar neutrino
flux and the neutrino oscillation parameters $\Delta m^2$ and $\tan^2\theta_{12}$. In
table~\ref{tab:rangeSplusK}, we follow the usual convention and represent the $^8$B solar neutrino flux
by the parameter $f_{\rm B}$, where $f_{\rm B}$ is the ratio of the true flux to the flux predicted by
the standard solar model as defined in eq.~\ref{eq:fbtotaldefinition}.

The comparison of the first and second rows of table~\ref{tab:rangeSplusK} shows the impact of the recent
SNO~\cite{snosalt}, GNO~\cite{gno}, and SAGE~\cite{sage02} data, especially the  recently released SNO
salt phase data~\cite{snosalt,howto2}. The positive $3\sigma$ uncertainty in $\Delta m^2$ is decreased by
a factor of 4.5. The solar data before (after) the SNO salt phase release disfavors the higher LMA
island less (more) than $\Delta \chi^2 = 9$ . The SNO salt phase data also improved significantly our
knowledge of $\tan^2\theta_{12}$, especially the higher values;  the $3\sigma$ uncertainty in
$\tan^2\theta_{12}$ is decreased by a factor of order 1.5. Moreover, the $1\sigma$ and $3\sigma$
uncertainties in $f_{\rm B}$ are decreased by a 25\%.

 The $1\sigma$ ranges of $\Delta m^2$ and $\tan^2\theta_{12}$ are slightly
different in figure~\ref{fig:SplusK} and in table~\ref{tab:rangeSplusK}. Figure~\ref{fig:SplusK} was
constructed by marginalizing over $\theta_{13}$ and $f_{\rm B}$; the confidence level for the figure was
calculated for two degrees of freedom. In constructing each column of table~\ref{tab:rangeSplusK}, we
marginalized over all of the variables except the one whose range is listed; the confidence levels for
the table were calculated for one degree of freedom.

The longer operation of the KamLAND experiment will shrink significantly the uncertainties in $\Delta
m^2$. The $1\sigma$ ($3\sigma$) uncertainty in $\Delta m^2$ is expected to
decrease by a factor of order 2 (2.5). The expected improvements in our knowledge of
$\tan^2\theta_{12}$  and of  the $^8$B neutrino flux are  much more modest.

\subsection{$^7$Be and $^8$B neutrino fluxes as free parameters}
\label{subsec:SplusKbe7b8}

Table~\ref{tab:rangeSplusKb8be7} shows how the allowed range of the oscillation parameters and the solar
neutrino fluxes change when the flux of $^7$Be solar neutrinos is treated as a free parameter.
Figure~\ref{fig:be7fig} shows the $1\sigma$ contours in the allowed oscillation parameter plane for the
case in which  the $^7$Be solar neutrino flux is treated as a free parameter (dashed curve) and also for
the case in which the $^7$Be neutrino flux is constrained to have the standard solar model best-fit value
and uncertainty (dotted curve). These two extreme treatments of the $^7$Be solar neutrino flux produce
very similar allowed regions in the $\Delta m^2$-$\tan^2\theta_{12}$ plane.

 The allowed ranges of $\Delta m^2$, $\tan^2\theta_{12}$, and $f_{\rm B}$ are only marginally affected by
letting the $^7$Be neutrino flux vary.

However, the first row of table~\ref{tab:rangeSplusKb8be7} shows that the current $1\sigma$ experimental
uncertainty in the reduced $^7$Be neutrino flux, $f_{\rm Be}$ (see eq.~\ref{eq:fbedefinition}), is a
factor of 2.3 times larger than the quoted theoretical uncertainty in the standard
solar model
calculation of the $^7$Be neutrino flux.  Moreover, table~\ref{tab:rangeSplusKb8be7} shows that the
experimental uncertainty in determining the $^7$Be solar neutrino flux  is not expected to decrease
significantly as a result of running the KamLAND experiment for three years. The KamLAND experiment does
not provide a strong constraint on the oscillation probability for neutrinos with energies comparable to
the $0.86$ MeV possessed by the $^7$Be solar neutrinos.

\TABLE[h!t]{\label{tab:rangeSplusKb8be7}\caption{{\bf Allowed neutrino parameters with free $^7$Be and
$^8$B solar neutrino fluxes: Solar plus KamLAND measurements.} The table presents the global best-fit and
$1\sigma$ ($3\sigma$) allowed ranges for the $^7$Be and $^8$B solar neutrino fluxes and for the neutrino
oscillation parameters $\Delta m^2$ and $\tan^2 \theta_{12}$. The allowed ranges are determined from a
global analysis, with one degree of freedom, of all currently available solar
(S)~~\cite{chlorine,sage02,gallex,superk,snoccnc,snodaynight,gno,snosalt} and KamLAND
(K)~\cite{kamlandfirstpaper} data. The $^7$Be and $^8$B solar neutrino fluxes are treated as free
parameters, but all other solar neutrino fluxes are assumed to have the standard solar model (BP00)
predicted values and uncertainties. In constructing each column of the table, we have marginalized over
all variables except the one whose range is shown.}
\begin{tabular}{lcccc}
\hline \noalign{\smallskip} Experiments & $\Delta m^2 (10^{-5}
eV^2)$&$\tan^2\theta_{12}$&$f_B$& $f_{Be}$ \\
\hline \noalign{\smallskip}\noalign{\smallskip} ~~~S + K & $7.3^{+0.4}_{-0.6}$ ($^{+7.7}_{-2.0}$) &
$0.40^{+0.06}_{-0.04}$ ($^{+0.23}_{-0.13}$) & $1.02^{+0.03}_{-0.05}$ ($^{+0.12}_{-0.14}$)
& $0.64^{+0.24}_{-0.22}$ ($^{+0.73}_{-0.64}$)\\
\noalign{\smallskip}\noalign{\smallskip} S + K 3 yr & $7.3^{+0.2}_{-0.2}$ ($^{+0.9}_{-0.6}$) &
$0.41^{+0.04}_{-0.04}$ ($^{+0.17}_{-0.11}$) & $1.01^{+0.04}_{-0.03}$ ($^{+0.10}_{-0.10}$) &
$0.66^{+0.22}_{-0.23}$ ($^{+0.68}_{-0.66}$)\\
\noalign{\smallskip} \hline
\end{tabular}
}

We have carried out global solutions with and without including the measurement~\cite{chlorine} of the
solar neutrino capture rate in chlorine (cf. table~\ref{tab:radiochemical}).  The purpose of these
calculations was to determine how sensitive to a single experiment are the current inferences regarding
neutrino oscillation parameters and solar neutrino fluxes.  In our view, the implications of solar
neutrino experiments are too important to rest solely on the results of a single measurement. Including
the chlorine experiment causes only a modest improvement in the allowed ranges, of order
 10\%, for
the neutrino oscillation parameters and for the $^8$B neutrino flux. However, the chlorine measurement
affects significantly the best-fit value for the $^7$Be neutrino flux and the inferred uncertainty in
this flux. With the chlorine experiment,
$f_{\rm Be} = 0.64^{+0.24}_{-0.22}$ ($^{+0.73}_{-0.64}$), while
without the chlorine measurement
$f_{\rm Be} = 0.97^{+0.29}_{-0.32}$ ($^{+0.92}_{-0.92}$).  The chlorine
measurement drives down the best-estimate for $f_{\rm Be}$ by about 30\% and
decreases the $1\sigma$ error on $f_{\rm Be}$ by almost 30\% .

\subsection{All neutrino fluxes as free parameters plus luminosity constraint}
\label{subsec:SplusKppbe7b8}

\TABLE[h!t]{\caption{{\bf Allowed neutrino parameters with free $p-p$, $^7$Be, and $^8$B solar neutrino
fluxes: with and without luminosity constraint.} Analysis A corresponds to a global analysis in which the
$p-p$, $^8$B, and $^7$Be solar neutrino fluxes were treated as free parameters. For Analysis B, the CNO
neutrino fluxes were also treated as free parameters as well as the $p-p$, $^8$B, and $^7$Be fluxes. The
symbol `+ lum' indicates that for rows two and three of table~\ref{tab:rangeSplusKppb8be7} the luminosity
constraint~\cite{luminosity} is included in the analysis. The table presents the global best fit and
$1\sigma$ ($3\sigma$) allowed ranges for the $p-p$, $^7$Be, and $^8$B solar neutrino fluxes and for the
neutrino mixing parameter $\tan^2\theta_{12}$. For all cases presented in this table, $\Delta m^2 =
7.3^{+0.4}_{-0.6}\times 10^{-5} {\rm eV^2} $ The results given here were obtained using all the currently
available data from the solar~~\cite{chlorine,sage02,gallex,superk,snoccnc,snodaynight,gno,snosalt} and
KamLAND~\cite{kamlandfirstpaper} neutrino experiments.  All other (much less important) solar neutrino
fluxes are assumed to have the standard solar model (BP00) predicted values and uncertainties. In
constructing each column of the table, we have marginalized over all variables except the one whose range
is shown in the column of interest. \label{tab:rangeSplusKppb8be7}}
\begin{tabular}{lcccc}
\hline \noalign{\smallskip} Analysis &$\tan^2\theta_{12}$&$f_B$& $f_{Be}$ & $f_{pp}$ \\
\hline \noalign{\smallskip}\noalign{\smallskip} A & $0.45^{+0.04}_{-0.06}$ ($^{+0.24}_{-0.16}$) &
$0.99^{+0.05}_{-0.03}$ ($^{+0.14}_{-0.13}$) & $0.13^{+0.41}_{-0.13}$ ($^{+1.27}_{-0.13}$)
& $1.38^{+0.18}_{-0.25}$ ($^{+0.47}_{-0.75}$)\\
\noalign{\smallskip}\noalign{\smallskip} A + lum & $0.40^{+0.06}_{-0.04}$ ($^{+0.23}_{-0.12}$) &
$1.02^{+0.03}_{-0.05}$ ($^{+0.12}_{-0.14}$) & $0.58^{+0.26}_{-0.25}$ ($^{+0.81}_{-0.58}$)
& $1.03^{+0.02}_{-0.02}$ ($^{+0.05}_{-0.06}$)\\
\noalign{\smallskip}\noalign{\smallskip} B + lum & $0.41^{+0.05}_{-0.05}$ ($^{+0.22}_{-0.13}$) &
$1.01^{+0.04}_{-0.04}$ ($^{+0.13}_{-0.13}$)& $0.93^{+0.25}_{-0.63}$ ($^{+0.80}_{-0.93}$)
& $1.02^{+0.02}_{-0.02}$ ($^{+0.06}_{-0.06}$)\\
\noalign{\smallskip} \hline
\end{tabular}
}

Table~\ref{tab:rangeSplusKppb8be7} presents the results of global analyses in which the $p-p$ solar
neutrino flux, as well as the $^7$Be and $^8$B solar neutrino fluxes, are treated as free parameters.
This is the first table in which we show results obtained with and without imposing the luminosity
constraint~\cite{luminosity}.  Cases in which the luminosity constraint are included are denoted by `+
lum' in the table.

Figure~\ref{fig:beforeafter} illustrates (primarily) the impact of the SNO salt-phase data~\cite{snosalt}
announced at the TAUP03 conference on September 7, 2003. The left panel of the figure shows the allowed
oscillation regions (at  $90$\%, $95$\%, $99$\%, and $99.73$\% ($3\sigma$) ) before the salt-phase data
were announced and the right panel shows the currently allowed oscillation regions.
 \FIGURE[!t]{ \centerline{\psfig{figure=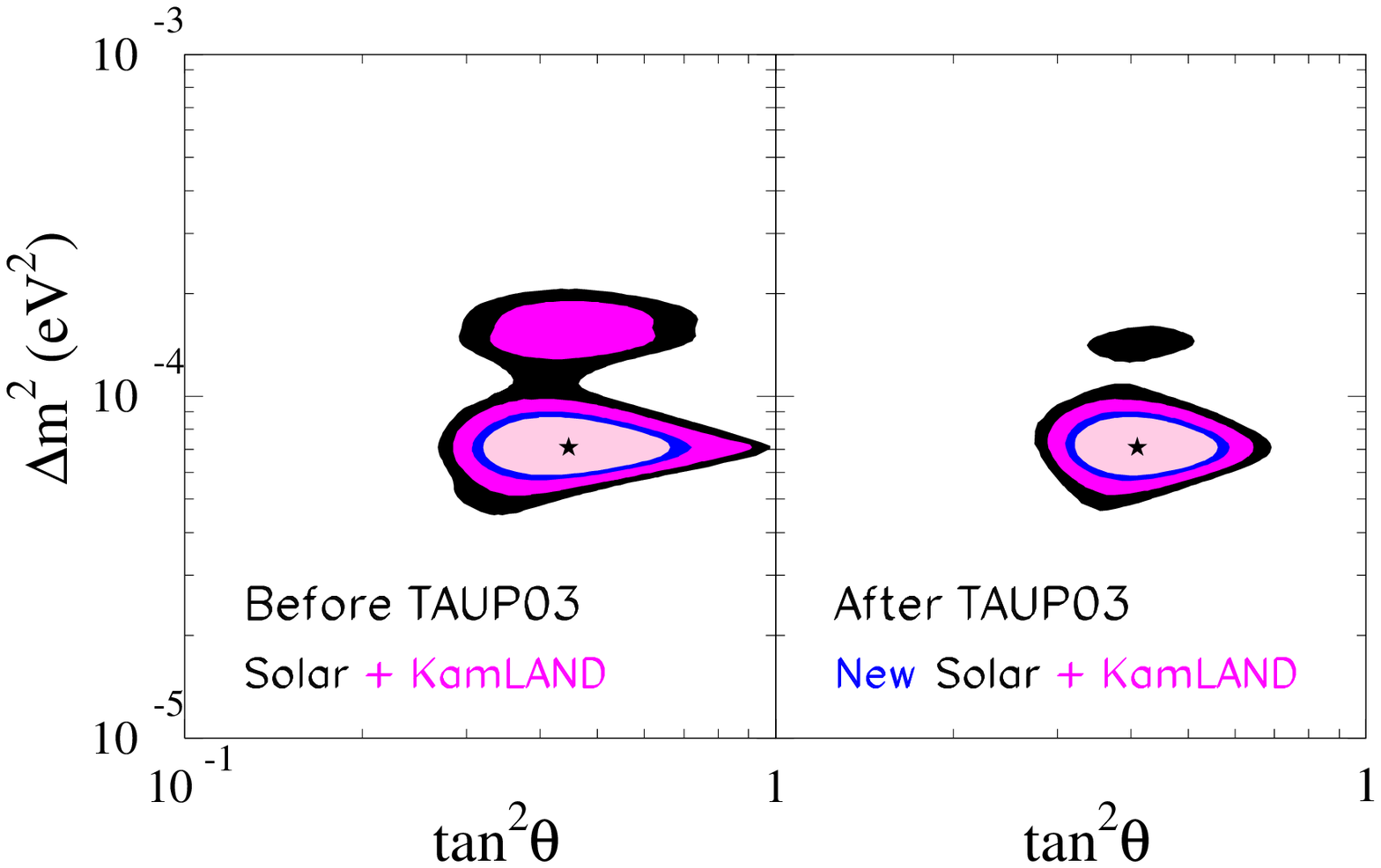,width=4.5in}}\caption{{\bf ``Before and
 After TAUP03.''} The left panel shows the allowed regions for neutrino
oscillations computed using all solar neutrino and reactor experimental data available prior to September
07, 2003. The right panel shows the allowed regions computed with the same procedures but including the
SNO salt-phase data~\cite{snosalt} and the improved SAGE~\cite{sage02} and GALLEX-GNO~\cite{gno} data
that were announced at the conference TAUP03 on September 7, 2003. The CL contours shown in the figure
are $90$\%, $95$\%, $99$\%, and $99.73$\% ($3\sigma$). The global best-fit points are marked by a star.
\label{fig:beforeafter}}}

\subsubsection{Neutrino fluxes without the luminosity constraint}
\label{subsubsec:fluxeswithout}

The first row of table~\ref{tab:rangeSplusKppb8be7}, which was calculated without including the
luminosity constraint, should be compared with the second row
of table~\ref{tab:rangeSplusK} and the first row of table~\ref{tab:rangeSplusKb8be7}.
The accuracy with which the global solution appears to determine the
solar neutrino fluxes degrades as more free neutrino fluxes are varied in the analysis. For example, the
quoted BP00 error on the predicted $^7$Be neutrino flux is $\pm 10$\% ($1\sigma$), whereas the
experimental determination is only accurate to $\pm 26$\% when the $^7$Be neutrino
flux is treated as a
free parameter. Similarly, the SSM uncertainty on the predicted $p-p$ neutrino flux is $\pm 1$\%, but the
flux is only determined experimentally to $\pm 21$\%. The $^8$B solar neutrino flux,
which is primarily
determined by the SNO and Super-Kamiokande measurements, is not affected much by treating additional
neutrino fluxes as free parameters.  The accuracy with which
$\Delta m^2$ and $\tan^2 \theta_{12}$ are determined is also not
affected significantly by adding more free fluxes.

The luminosity constraint is grossly violated if we allow the $p-p$, $^7$Be, and $^8$B neutrino fluxes to
vary as free parameters without requiring that the total luminosity equal  the observed luminosity. The
first row of table~\ref{tab:rangeSplusKppb8be7} shows that the best-fit value for the $p-p$ neutrino flux
is
$f_{\rm p-p} = 1.38$ in this case, whereas the maximum value allowed by the luminosity constraint is
$f_{\rm p-p} = 1.09$ (see equation~\ref{eq:approxluminosity}).

\subsubsection{Neutrino fluxes with the luminosity constraint}
\label{subsubsec:fluxeswith}

The constraints on the $p-p$ neutrino flux are dramatically improved, by more than a factor of ten, when
the luminosity constraint is applied.  This improvement is independent of whether the CNO neutrino fluxes
are assumed to have their standard solar model best-fit and uncertainties (row two of
table~\ref{tab:rangeSplusKppb8be7}) or whether the CNO fluxes are treated as free parameters (row three
of table~\ref{tab:rangeSplusKppb8be7}). Remarkably, the $p-p$ solar neutrino flux is known at present to
$\pm 2$\% (1$\sigma$ uncertainty) and is in agreement to that accuracy with the prediction of the
standard solar model~\footnote{The results given here should be compared with an insightful analysis
given in section~VI of ref.~\cite{sage02}. The authors of ref.~\cite{sage02} made a series of simplifying
assumptions in analyzing the pre-KamLAND solar neutrino data and obtained $f_{p-p} = 1.29 \pm 0.30$. The
principal assumptions made in this analysis, which did not include a global treatment of all the
experiments together, included assuming the correctness of the BP00 fluxes for the $^7$Be, CNO, and $pep$
neutrinos and the constancy of the survival probability for all of these neutrinos.  Although the authors
suggest that their uncertainty may be underestimated, their reasoning shows clearly how the gallium
experiments constrain the $p-p$ neutrino flux.}

The constraints on the $^7$Be neutrino flux are not improved by imposing the luminosity constraint. The
luminosity constraint affects differently the $p-p$ and $^7$Be solar neutrino fluxes. We shall explain
the origin of this difference in section~\ref{sec:luminosityexplain}.  However, the alert reader may
notice a counter-intuitive result shown in column~4 ($f_{\rm Be}$) of table~\ref{tab:rangeSplusKppb8be7}.
The uncertainties are actually larger for $f_{\rm Be}$ when the luminosity constraint is imposed. The
reason for this anomalous behavior is that the global $\chi^2_{\rm min}$ is smaller
($\chi^2_{\rm min} =
78.6$) for case A of table~\ref{tab:rangeSplusKppb8be7} (without the luminosity constraint) than  the
global  $\chi^2_{\rm min}$ (80.4) found for case A plus the luminosity constraint.
The uncertainties at a
fixed confidence level ($\Delta \chi^2$) are calculated by computing the range of $f_{\rm Be}$ allowed by
the relation $\Delta \chi^2 = \chi^2(f_{\rm Be}) - \chi^2_{\rm min}$.

The CNO fluxes are  poorly constrained by the available solar neutrino data (cf. ref.~\cite{cnopaper}).
In all cases, that we have considered the favored solution has the $^{13}$N, $^{15}$O, and $^{17}$F solar
neutrino fluxes all equal to zero. The $^{15}$O flux is the best determined of the CNO solar neutrino
fluxes. We find in units of the standard solar model prediction~\cite{bp00} that

\begin{equation}
f_{\rm O}=0.0^{+2.4}_{-0.0}(^{+6.4}_{-0.0})\, . \label{eq:o15pluslum}
\end{equation}
The constraint given in equation~\ref{eq:o15pluslum} is unaffected, to the accuracy shown, by whether or
not the luminosity constraint is imposed.

At $1\sigma$, the $^{13}$N solar neutrino flux could be as large as $13$ times the standard solar model
prediction and the $^{17}$F neutrino flux could be as large as 220 times the solar model prediction.

\subsubsection{Allowed oscillation solutions: with and without the luminosity constraint}
\label{subsubsec:allowedoscillations}

In ref.~\cite{postkamland}, the authors found that, using the existing solar plus KamLAND data, only the
LMA oscillation solution is allowed at 4.7 $\sigma$. Repeating the same analysis but including the SNO
salt phase data and the other new solar neutrino data~\cite{sage02,snosalt,gno}, we find that only the
LMA oscillation solution is allowed at 5.2 $\sigma$.

 All other solar neutrino oscillation solutions (e.g., LOW, VAC, SMA) are excluded. In the analysis described in ref.~\cite{postkamland} and in other
similar discussions (see, e.g., ref.~\cite{analysispostkamland}), the solar neutrino fluxes and their
uncertainties were either all taken from the standard solar model or all of the fluxes except the $^8$B
neutrino flux were taken from the standard solar model.

 If, instead,  all solar neutrino fluxes are regarded as free parameters and the luminosity
constraint is imposed, we find that the LMA oscillation is the only allowed solution at
 4.5 $\sigma$.

All solutions in which the neutrino energy spectra are undistorted, i.e., all solutions without some
form of neutrino oscillations (what used to be called `non-standard solar models'), are disfavored at
 $8.3\sigma$~\cite{luminosity}.

The existing constraints on the $^7$Be and $p-p$ neutrino fluxes result from the chlorine~\cite{chlorine}
and gallium~\cite{sage02,gallex,gno} solar neutrino experiments (cf. table~\ref{tab:radiochemical}). It
is of interest to test separately the sensitivity of the inferred constraints on the neutrino fluxes by
omitting either the chlorine or the gallium experiments from the global analysis (but imposing the
luminosity constraint as in the third row of table~\ref{tab:rangeSplusKppb8be7}). Omitting the chlorine
experiment, we find that the constraints on $\Delta m^2$, $\tan^2 \theta_{12}$, and $f_{\rm B}$ are
essentially as stated in the third row of table~\ref{tab:rangeSplusKppb8be7}, but that the constraints on
the $^7$Be and $p-p$ neutrino fluxes are much relaxed:

\begin{equation}
{\rm Without ~Cl:}~~f_{\rm Be}~=~1.24^{+0.31}_{-1.24} \left(^{+0.76}_{-1.24}\right);~~ f_{p-p}~=~1.00^{+0.04}_{-0.02} (^{+0.09}_{-0.08}).
 \label{eq:relaxcl}
\end{equation}
Omitting the gallium experiment, we find
\begin{equation}
{\rm Without ~Ga:}~~f_{\rm Be}~=~0.33^{+0.64}_{-0.33} \left(^{+1.43}_{-0.33}\right);~~ f_{p-p}~=~1.05^{+0.02}_{-0.04} (^{+0.04}_{-0.10}).
 \label{eq:relaxga}
\end{equation}

The chlorine experiment has a stronger effect on the constraints than does the gallium experiment.

Figure~\ref{fig:SplusK} shows that the  uncertainties in determining  $\Delta m^2$ and  $\tan^2
\theta_{12}$ from existing data are not affected significantly by whether only the $^8$B neutrino flux is
treated as a free variable or whether, instead, all the fluxes are treated as free variables subject to
the luminosity constraint. In both cases, the allowed ranges are practically the same (cf. the  second row
of table~\ref{tab:rangeSplusK} with the last row of table~\ref{tab:rangeSplusKppb8be7}).

\subsection{Sterile neutrinos}
\label{subsec:sterile}

We can parameterize the sterile contribution to the neutrino flux in terms of  $\sin^2\eta$ or,
alternatively, in terms of a derived parameter $f_{\rm B,\,sterile}$, the sterile fraction of the $^8$B
neutrino flux (see discussion in section~\ref{subsec:globalchi} and the more complete description in
ref~\cite{postkamland}).

The 1$\sigma$ allowed range for the active-sterile admixture is
\begin{equation}
\sin^2\eta\leq 0.09
\label{eq:etacurrent}
\end{equation}
 in our analysis of the existing solar plus KamLAND data (cf. the first row of
table~\ref{tab:rangeSplusK}). The fundamental parameter describing the sterile fraction is $\sin^2\eta$
(cf. section~\ref{subsec:globalchi} and ref~\cite{postkamland}).  However, it is convenient to think in
terms of a sterile fraction of the flux, $f_{\rm B,\, sterile}$ , which is potentially observable in the
Super-Kamiokande and SNO experiments. This range corresponds to
\begin{equation}
f_{\rm B,\, sterile}~=~0.0^{+0.06}_{-0.00}~~({\rm solar + KamLAND})\,. \label{eq:fbsterileK}
\end{equation}
 Three years of accumulation of data by KamLAND (the S + K 3
yr second row of table~\ref{tab:rangeSplusK}) can marginally improve
this bound to
\begin{equation}
\sin^2\eta\leq 0.08 \, ,
\label{eq:eta3years}
\end{equation}
or
\begin{equation}
f_{\rm B,\, sterile}~=~0.0^{+0.06}_{-0.00}~~({\rm solar + KamLAND~3~yr})\,. \label{eq:fbsterileK3yr}
\end{equation}

On the other hand, if we impose the luminosity constraint while allowing the $p-p$, $^7$Be, $^8$B, and
CNO neutrino fluxes to vary freely (analysis B + lum in table~\ref{tab:rangeSplusKppb8be7}), our present
bounds are slightly weaker. In this case, we find
\begin{equation}
\sin^2\eta\leq 0.10 \, ,
\label{eq:etalc}
\end{equation}
 or
\begin{equation}
f_{\rm B,\, sterile}~=~0.0^{+0.07}_{-0.00}~~({\rm solar + KamLAND; \,\, free~fluxes;
luminosity~constraint})\,. \label{eq:fbsterileKfree}
\end{equation}

\subsection{CNO luminosity}
 \label{subsec:cno}
How much of the solar luminosity is due to the carbon-nitrogen-oxygen (CNO) nuclear fusion reactions that
Hans Bethe suggested in 1939 are primarily responsible for energy generation in the Sun?

In ref.~\cite{cnopaper}, the authors derived $1\sigma$ ($3\sigma$) constraints on the allowed range of
the CNO contribution to the solar luminosity using all the presently existing solar neutrino and KamLAND
experimental data.   Imposing the luminosity constraint and carrying out a global analysis of the ratio
of the CNO-produced luminosity to the total luminosity, they found~\cite{cnopaper}
\begin{equation}
\frac{L_{\rm CNO}}{L_\odot} ~=~ 0.0_{-0.0}^{+2.8}\,\% ~(0.0_{-0.0}^{+7.3}\,\%)~~{\rm at ~1\sigma
\,(3\sigma)} ~~{(\rm with~luminosity~constraint)}\,. \label{eq:cnolimitlc}
\end{equation}
We have carried out a similar global analysis but using simulated three years of data for KamLAND. The
results summarized in equation~\ref{eq:cnolimitlc} are not affected, to the accuracy shown, by the
additional simulated KamLAND data.

For completeness, we have also carried out a global analysis allowing the fluxes to be free but without
imposing the luminosity constraint. We find
\begin{equation}
\frac{L_{\rm CNO}}{L_\odot} ~=~ 1.6_{-1.0}^{+1.4}\,\% ~(1.6_{-1.0}^{+5.2}\,\%)~~{\rm at ~1\sigma
\,(3\sigma)} ~~{(\rm no~luminosity~constraint)}\,. \label{eq:cnolimitnolc}
\end{equation}

 The best-estimate values given in equation~\ref{eq:cnolimitlc} and equation~\ref{eq:cnolimitnolc} are
different. If the luminosity constraint is not imposed, then the global solution prefers a non-zero value
for the CNO luminosity that is just slightly larger than the standard solar model best-estimate of 1.5\%.
The low-energy neutrinos that are required to fit the radiochemical experimental data are provided in
this case by the CNO reactions. Moreover, the chlorine and gallium experiments  suggest a preferred value
of zero for the $^7$Be solar neutrino flux (see case A of table~\ref{tab:rangeSplusKppb8be7}).

However, if the luminosity constraint is imposed, then the $^7$Be solar neutrino flux cannot easily be
zero and the required low energy neutrinos are produced by $^7$Be electron capture (see case `B + lum' of
table~\ref{tab:rangeSplusKppb8be7} and the approximate form of the luminosity constraint,
equation~\ref{eq:approxluminosity}). In this case, the radiochemical experiments prefer a zero-value for
the CNO luminosity.

The upper limits given in equations~\ref{eq:cnolimitlc} and \ref{eq:cnolimitnolc} are relatively
insensitive to whether or not the luminosity constraint is imposed.

\subsection{Total luminosity measurement with neutrinos} \label{subsec:totalluminosity}

What limits do existing solar plus reactor neutrino experiments place on the total luminosity of the Sun?

 Each solar neutrino flux, $\Phi_i$, corresponds to a well-defined flux of energy, $\alpha_i
 \Phi_i$, where the energy coefficients are given in table~1 of  ref.~\cite{luminosity}. This energy is,
 of course, much larger than the actual energy of the neutrinos since most of the energy from solar fusion
 reactions is released to the star in the form of thermal energy and eventually radiated away in
 the form of photons (see discussion in ref.~\cite{luminosity}).

 We have performed a global analysis of all the available solar and reactor data to find the $1\sigma$
 ($3\sigma$) allowed range for the total flux of solar luminosity, $\sum_i \alpha_i \Phi_i$, based solely
 upon the neutrino data. We find for the ratio of the neutrino-inferred solar luminosity, $L_\odot{\rm
  (neutrino-inferred)}$, to the accurately measured photon luminosity, $L_\odot$, that

\begin{equation}
 \frac{L_\odot{\rm
  (neutrino-inferred)}}{L_\odot}~=~1.4^{+0.2}_{-0.3} \left(^{+0.7}_{-0.6}\right).
 \label{eq:lnuoverlphoton}
\end{equation}
At $3\sigma$, the neutrino-inferred solar luminosity can be as large as (as small as) 2.1 (0.8) the
precisely measured photon-luminosity.  Three years of KamLAND data is expected to improve the limits
stated in equation~\ref{eq:lnuoverlphoton} by only about 10\%.

\subsection{Predictions for the radiochemical chlorine and gallium experiments}
\label{subsec:predictclga}

 Table~\ref{tab:radiochemical} summarizes the predictions for the radiochemical
solar neutrino experiments of the LMA solution that are inferred  from the existing solar neutrino and
KamLAND experiments. The predictions are based upon the global analyses for which the best-fit neutrino
oscillation parameters are listed in the first row of table~\ref{tab:rangeSplusK} (and the third row of
table~\ref{tab:rangeSplusKppb8be7}, free solar neutrino fluxes plus luminosity constraint). The measured
rate in the chlorine experiment is approximately $2\sigma$ ($1\sigma$) smaller than the best-fit LMA
prediction.

\begin{table}[!t]
\caption{{\bf Neutrino oscillation predictions for the chlorine and gallium radiochemical experiments.}
The rates without parentheses (in parentheses) are presented for the best-fit oscillation parameters of
the allowed solutions listed in table \ref{tab:rangeSplusK} ,which assumes neutrino fluxes from the BP00
solar model  except for the $^8$B neutrino flux (table~\ref{tab:rangeSplusKppb8be7}, which was derived
with free neutrino fluxes and the luminosity constraint). The predictions are based upon the global
analysis of existing solar plus KamLAND data and  the neutrino absorption cross sections given in
refs.~\cite{neutrinoastrophysics,chlorineb8cs,GaCS}.   The total rates should be compared with the
standard solar model values \cite{bp00}, which are $7.6^{+1.3}_{-1.1}$ SNU (chlorine) and $128^{+9}_{-7}$
SNU (gallium), and the measured values, which are $2.56 \pm 0.23$ (chlorine \cite{chlorine}) and $69.2
\pm 4.0$ (gallium \cite{sage02,gallex,gno}). \label{tab:radiochemical} }
\begin{center}
\begin{tabular}{@{\extracolsep{5pt}}lcc}
\hline \noalign{\smallskip}
Source&Cl&Ga\\
&(SNU)&(SNU)\\
&LMA&LMA\\
\noalign{\smallskip} \hline \noalign{\smallskip}
$p-p$            &  0(0)   & 40.6 (41.8)\\
$pep$           & 0.13(0.13) & 1.57 (1.61)\\
$hep$           & 0.02(0.02) & 0.03 (0.03)\\
${\rm ^7Be}$  & 0.64(0.60) & 19.0 (17.9)\\
${\rm ^8B}$   & 2.02(2.00) & 4.28 (4.23)\\
${\rm ^{13}N}$& 0.05(0.00) & 1.83 (0.00)\\
${\rm ^{15}O}$& 0.18(0.00) & 2.94 (0.00)\\
${\rm ^{17}F}$& 0.00(0.00) & 0.03 (0.00)\\
\noalign{\medskip}
&\hrulefill&\hrulefill\\
Total         & $3.04\pm 0.04 (2.75\pm 0.18) $ &
$70.3^{+2.6}_{-2.3}(65.6^{+4.2}_{-4.8})$ \\
\noalign{\smallskip} \hline
\end{tabular}
\end{center}
\end{table}

\section{How does the luminosity constraint affect the allowed range of solar neutrino fluxes?}
\label{sec:luminosityexplain}

The luminosity constraint states that a specific linear combination of all the neutrino fluxes that are
produced by fusion reactions  at a temperature of order 1 keV must be equal to unity.  The precise
coefficients for this linear combination are given in table~1 and equation~2 of ref.~\cite{luminosity}.

We can understand the strong effect of the luminosity constraint on the $p-p$ neutrino flux, and the
relatively weak effect on the $^7$Be neutrino flux, by including only the largest terms in the luminosity
constraint. We define a reduced neutrino flux  $\phi$ to be the ratio of the true solar neutrino flux
from a particular nuclear reaction to the neutrino flux predicted for this source by the BP00 solar
model. If the standard solar model is a relatively good approximation to the actual Sun, then the leading
terms in the luminosity constraint are
\begin{equation}
\phi(p-p) ~\approx~ 1.09 ~-~ 0.08\phi(^7{\rm Be}) ~-~0.01\phi(^{15}{\rm O}).
 \label{eq:approxluminosity}
\end{equation}
We see immediately from equation~\ref{eq:approxluminosity}  that the luminosity constraint does not allow
the $p-p$ flux to exceed, at any confidence level, 9\% of the current standard solar model predicted
flux. By contrast, if we do not impose the luminosity constraint, the best-fit value for $\phi(p-p)$ is
1.49 not 1.0 and the maximum value of  $\phi(p-p)$ is 2.02 at $3\sigma$ (see the last column of
table~\ref{tab:rangeSplusKppb8be7}). The strong lower limit on $\phi(p-p)$ follows from
equation~\ref{eq:approxluminosity} because the chlorine and gallium solar neutrino experiments imply that
$\phi(^7{\rm Be}) $ can not be much larger than unity (cf. table~\ref{tab:radiochemical}).

The small coefficient of $\phi(^7{\rm Be})$ in equation~\ref{eq:approxluminosity} is the reason that the
luminosity constraint does not impose tight constraints on the allowed $^7$Be solar neutrino flux,
provided that the standard solar model is a reasonable approximation to the actual Sun. In principle, the
$^7$Be solar neutrino flux could be as large as three times the predicted BP00 $^7$Be neutrino flux
without violating nuclear physics or energy constraints~\cite{luminosity}. However, this would be true
only for a real Sun that is very different from the standard solar model.

\section{What will we learn from a $\mathbf{^7}$Be solar neutrino experiment?}
\label{sec:whatbe7}

In this section, we investigate quantitatively what we may expect to learn from a measurement of the
$^7$Be solar neutrino flux.

We begin in section~\ref{subsec:be7implications} by comparing the differences in the predictions for
future experiments caused by treating the $^7$Be and other solar neutrino fluxes as free parameters
rather than by taking the fluxes, and their associated uncertainties, from the standard solar model. Then
in section~\ref{subsec:how7be} we investigate how accurately a measurement of the $^7$Be solar neutrino
flux is expected to determine neutrino oscillation parameters. We study in
section~\ref{subsec:howbefluxes} how accurately a $^7$Be experiment will determine solar neutrino fluxes.
We determine in section~\ref{subsec:cnoluminositybe} how well a $^7$Be neutrino-electron scattering
experiment can constrain the CNO-produced luminosity of the Sun and in
section~\ref{subsec:totalluminositybe} we show how accurately a $^7$Be neutrino-electron scattering
experiment will help measure the total luminosity of the Sun. We compare in section~\ref{subsec:cc7Be}
what can be learned from a $^7$Be CC experiment with what can be learned from a $\nu-e$ scattering
experiment.

 In all cases, we treat the $^8$B solar neutrino
flux as a free parameter.

\subsection{$^7$Be free: Implications for future experiments}
\label{subsec:be7implications}

Whether or not $f_{\rm Be}$ is treated as a free parameter  affects strongly the uncertainty, and the
best estimate, of the predicted rate at which electrons scatter $^7$Be neutrinos or are absorbed in a CC
experiment. We define the reduced $^7$Be $\nu-e$ scattering or CC (absorption) rate as follows:

\begin{equation}
[{\rm ^7Be}] ~\equiv~ \frac{\rm Observed ~ rate}{\rm BP00~ predicted ~rate}~, \label{eq:be7reducedrate}
\end{equation}
where the denominator of equation~\ref{eq:be7reducedrate} is the rate calculated with the BP00 $^7$Be
neutrino flux  assuming no neutrino oscillations.

If we perform the global analysis of solar plus reactor experiments assuming that the BP00 calculation
for the $^7$Be solar neutrino flux and its uncertainty are valid, then the  predicted rate in a $^7$Be
$\nu-e$ scattering experiment is, with $1\sigma$ ($3\sigma$) uncertainties:

\begin{equation}
\left[{\rm ^7Be}\right]_{\rm \nu-e}~=~ 0.66\pm 0.02\, (^{+0.05}_{-0.04})\,.
 \label{eq:be7esprediction}
\end{equation}
Thus, if we use the solar model calculation of the $^7$Be neutrino flux and its uncertainty, then the
predicted event rate for the $^7$Be rate experiment has a precision of $\pm 3$\%. The corresponding
prediction for the reduced event rate in a CC (absorption) experiment with $^7$Be solar neutrinos is:

\begin{equation}
\left[{\rm ^7Be}\right]_{\rm CC}~=~ 0.57 \pm 0.02 \,( ^{+0.07}_{-0.06}) \,.
 \label{eq:be7ccprediction}
\end{equation}
The precision of the CC prediction is comparable to what is obtained for a $\nu-e$ scattering experiment.
The results given in equation~\ref{eq:be7esprediction} and equation~\ref{eq:be7ccprediction} only make
use of the currently available KamLAND data, but the results are practically the same if we use the three
year simulated KamLAND result.

 The result shown in
equation~\ref{eq:be7ccprediction} for the CC reaction  includes  only $^7$Be neutrinos from the 0.86 MeV
electron capture line. On the other hand, the result for $\nu-e$ scattering that is given in
equation~\ref{eq:be7esprediction} includes neutrinos from all the solar neutrino fluxes that produce
recoil electrons with energies in the  range  0.25-0.8 MeV. Most of the recoil electrons in the selected
energy range are produced by $^7$Be neutrinos, although there are small contributions from CNO, $pep$, and
$p-p$ neutrinos (whose fluxes are taken from the BP00 solar model~\cite{bp00}).

The situation is very different if we treat the $^7$Be solar neutrino flux as a free variable in the
global analysis. In this case, using the existing KamLAND data and all the available solar data, we find:

\begin{equation}
\left[{\rm ^7Be}\right]_{\rm \nu-e}~=~ 0.47^{+0.13}_{-0.12}  (^{+0.39}_{-0.41})\,\, {\rm [free~ ^7Be
~flux]}\, ,
 \label{eq:be7freeesprediction}
\end{equation}
and

\begin{equation}
\left[{\rm ^7Be}\right]_{\rm CC}~=~ 0.38^{+0.14}_{-0.13} (^{+0.41}_{-0.38}) \,\, {\rm [free~ ^7Be
~flux]}\,.
 \label{eq:be7freeccprediction}
\end{equation}
The reason that the $3\sigma$ lower limit in equation~\ref{eq:be7freeesprediction} and
equation~\ref{eq:be7freeccprediction} does not reach zero is that the CNO contribution between 0.2 MeV
and 0.8 MeV is included, by definition, in $\left[{\rm ^7Be}\right]$.

The most important case allows all the neutrino fluxes to vary freely within the limits imposed by the
luminosity constraint. With this  experimental approach (cf. the third row of
table~\ref{tab:rangeSplusKppb8be7}), we find an even larger allowed range for the predicted rate of a
$^7$Be solar neutrino experiment. For this `all free' case, we obtain

\begin{equation}
\left[{\rm ^7Be}\right]_{\rm \nu-e}~=~ 0.58^{+0.14}_{-0.23}  (^{+0.45}_{-0.53})\,\, {\rm [all~free]}\, ,
 \label{eq:be7allfreeesprediction}
\end{equation}

\begin{equation}
\left[{\rm ^7Be}\right]_{\rm CC}~=~ 0.54^{+0.15}_{-0.30} (^{+0.41}_{-0.54}) \,\, {\rm [all~free]}\,.
 \label{eq:be7allfreeccprediction}
\end{equation}

 The uncertainties are enormous for the predicted rates in $^7$Be experiments (cf.
 equations~\ref{eq:be7allfreeesprediction}--\ref{eq:be7allfreeccprediction}) if the solar neutrino fluxes
 used in the global fits to experiments are unconstrained by
solar model calculations. The uncertainties in the predictions, $\sim \pm 60$\% of the best-estimate
value for the $\nu-e$ scattering rate ($\sim \pm 70$\% for the CC rate), are more than an order of
magnitude larger than the uncertainties in the predictions, $\sim \pm 3$\% for the $\nu-e$ scattering
rate ($\sim \pm 5$\% for the CC rate), if the $^7$Be neutrino flux is constrained by the standard solar
model prediction and uncertainties.

\subsection{How accurately will a $^7$Be solar neutrino experiment determine neutrino oscillation parameters?}
\label{subsec:how7be}

 We  suppose in this subsection that the $\nu-e$ scattering rate, $\left[{\rm ^7Be}\right]_{\rm \nu-e}$, is measured
 and that the best-fit value coincides with the result obtained assuming the standard solar
 model~\cite{bp00}
 $^7$Be neutrino flux and the preferred values for $\Delta m^2$ and $\tan^2 \theta_{12}$
 (cf. table~\ref{tab:rangeSplusK} and equation~\ref{eq:be7esprediction}). We have
then carried out a global solution including all the currently available solar neutrino data, the
simulated KamLAND three year data, and the hypothetical measurement of $\left[{\rm ^7Be}\right]_{\rm
\nu-e}$. We use the simulated three year data rather than the currently available one-year KamLAND data
since it seems likely that the KamLAND reactor experiment will be completed before a $^7$Be solar
neutrino experiment is completed.

We have imposed the luminosity constraint~\cite{luminosity} on the solar neutrino fluxes when carrying
out the global solution. We have treated the $p-p$, $^7$Be, $^8$B, and CNO neutrino fluxes as free
parameters.

 \TABLE[!t]{\caption{\label{tab:rangeslmanycases}{\bf The effect of a measurement of the
$^7$Be $\nu-e$ scattering rate.} The table presents best-fit and $1\sigma$ ($3\sigma$) ranges computed by
including all the currently available solar neutrino data
(S)~~\cite{chlorine,sage02,gallex,superk,snoccnc,snodaynight,gno,snosalt} plus three years of simulated
KamLAND measurements subject to the luminosity constraint. The $p-p$, $^7$Be, $^8$B, and CNO neutrino
fluxes are treated as free parameters. The last five rows of the table were computed using also a
simulated measurement of the reduced $\nu-e$ scattering rate (cf. equation~\ref{eq:be7reducedrate}),
$\left[{\rm ^7Be}\right]_{\rm \nu-e} = 0.66$, assumed to be measured with 30\%, 10\%, 5\%, 3\% or 1 \%
precision. In constructing each column of the table, we have marginalized over all variables except the
one whose range is shown. The allowed ranges of $\Delta m^2$ and $f_{\rm B}$ are, for all cases shown in
table~\ref{tab:rangeslmanycases}, $\Delta m^2 = 7.3^{+0.2}_{-0.2}\times 10^{-5} {\rm eV}^2$ and $f_{\rm
B} = 1.01^{+0.04}_{-0.04}$.  The allowed range of $\tan^2\theta_{12}$ changes only by a very small amount
as a result of a precise $^7$Be solar neutrino experiment.}
\begin{tabular}{lccc}
\hline \noalign{\smallskip} ~~~~Experiments &$\tan^2\theta_{12}$& $f_{Be}$
& $f_{pp}$ \\
\hline \noalign{\smallskip}\noalign{\smallskip} ~~~~~S + K 3 yr & $0.41^{+0.04}_{-0.04}$
($^{+0.15}_{-0.09}$) & $0.93^{+0.25}_{-0.63}$ ($^{+0.80}_{-0.93}$)
& $1.02^{+0.02}_{-0.02}$ ($^{+0.06}_{-0.06}$)\\
\noalign{\smallskip}\noalign{\smallskip} + $[^7$Be$]_{\nu-e} ~\pm 30\%$ & $0.41^{+0.04}_{-0.04}$
($^{+0.15}_{-0.09}$) & $0.99^{+0.20}_{-0.26}$ ($^{+0.61}_{-0.99}$)
& $1.015^{+0.015}_{-0.015}$ ($^{+0.047}_{-0.047}$)\\
\noalign{\smallskip}\noalign{\smallskip} + $[^7$Be$]_{\nu-e} ~\pm 10\%$ & $0.41^{+0.04}_{-0.04}$
($^{+0.15}_{-0.09}$) & $1.07^{+0.10}_{-0.11}$ ($^{+0.31}_{-0.45}$)
& $1.009^{+0.008}_{-0.008}$ ($^{+0.024}_{-0.028}$)\\
\noalign{\smallskip}\noalign{\smallskip} + $[^7$Be$]_{\nu-e} ~\pm 5\%$ & $0.41^{+0.04}_{-0.04}$
($^{+0.15}_{-0.09}$) & $1.09^{+0.05}_{-0.07}$ ($^{+0.17}_{-0.31}$)
& $1.008^{+0.005}_{-0.005}$ ($^{+0.013}_{-0.024}$)\\
\noalign{\smallskip}\noalign{\smallskip} + $[^7$Be$]_{\nu-e} ~\pm 3\%$  & $0.41^{+0.04}_{-0.04}$
($^{+0.15}_{-0.09}$) & $1.09^{+0.04}_{-0.05}$ ($^{+0.12}_{-0.26}$)
& $1.008^{+0.002}_{-0.005}$ ($^{+0.008}_{-0.024}$)\\
\noalign{\smallskip}\noalign{\smallskip} + $[^7$Be$]_{\nu-e} ~\pm 1\%$ & $0.41^{+0.04}_{-0.04}$
($^{+0.15}_{-0.09}$) & $1.10^{+0.02}_{-0.05}$ ($^{+0.07}_{-0.24}$)
& $1.007^{+0.002}_{-0.004}$ ($^{+0.005}_{-0.022}$)\\
\noalign{\smallskip} \hline
\end{tabular}
}

Table~\ref{tab:rangeslmanycases} compares the globally allowed ranges of  $\tan^2\theta_{12}$,  and $f_{\rm
Be}$  if a $^7$Be experiment is carried out with an accuracy of 30\%, 10\%, 5\%, 3\%, or 1\% . For
calibration, we also present in the first row of table~\ref{tab:rangeslmanycases} the allowed parameter
ranges in the absence of a $^7$Be experiment.

The allowed ranges of $\Delta m^2$ and $f_{\rm B}$ are, for all cases shown in
table~\ref{tab:rangeslmanycases}, $\Delta m^2 = 7.3^{+0.2}_{-0.2}\times 10^{-5} {\rm eV}^2$ and $f_{\rm
B} = 1.01^{+0.04}_{-0.05}$.  The allowed range of $\tan^2\theta_{12}$ changes only by a very small amount
as a result of a precise $^7$Be solar neutrino experiment.

\FIGURE[!t]{ \centerline{\psfig{figure=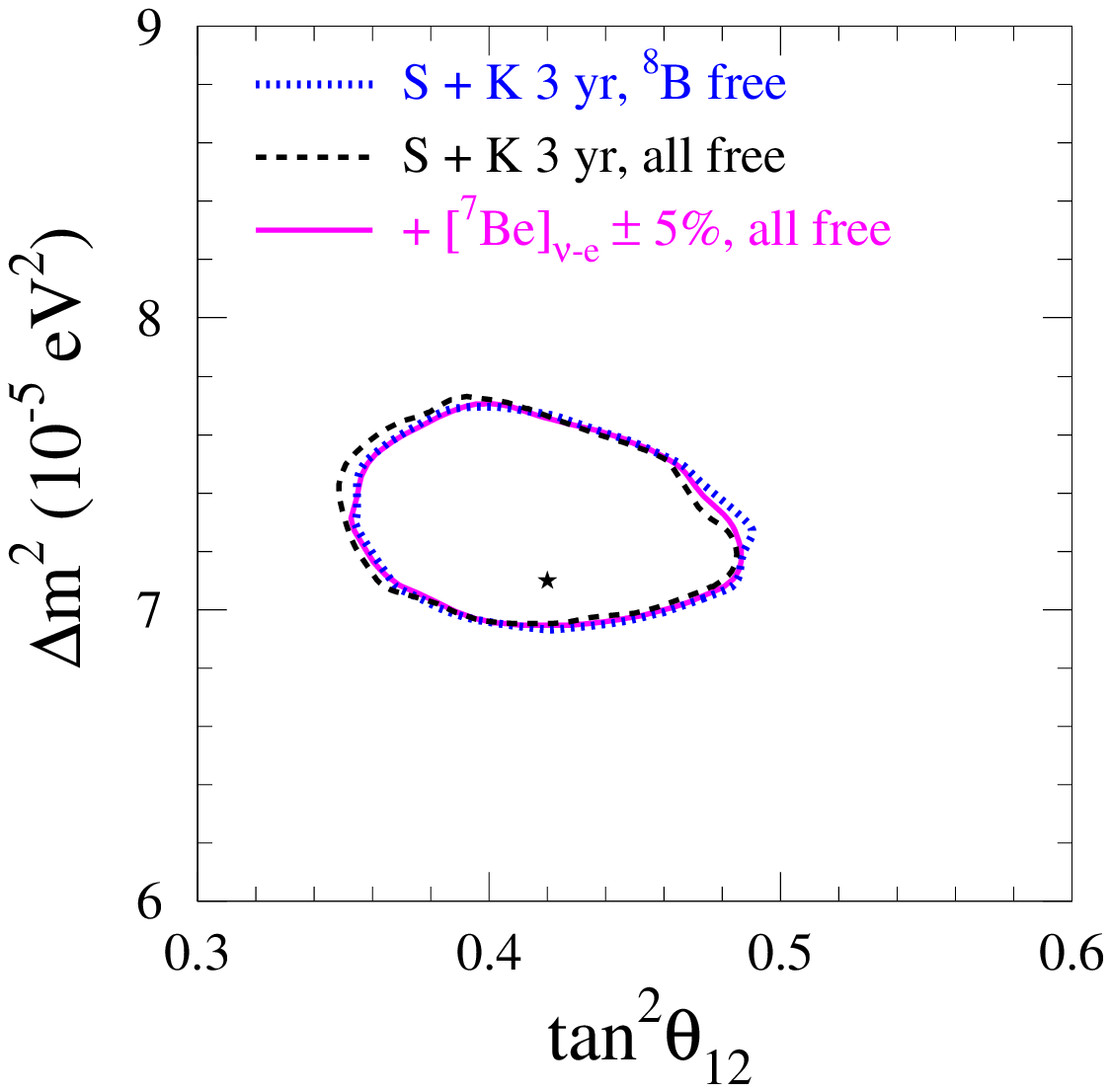,width=3.5in}} \caption{{\bf Effect of $^7$Be constraints
on the allowed regions for oscillation parameters.} The main result illustrated in this figure is that
the allowed region for oscillation parameters is not expected to be affected much by a measurement of the
$^7$Be solar neutrino flux. Each $1\sigma$ contour in the plane of allowed oscillation parameters is
constructed using all the currently available solar neutrino
data~~\cite{chlorine,sage02,gallex,superk,snoccnc,snodaynight,gno,snosalt} plus a simulated three years
of KamLAND data (cf. figure~\ref{fig:SplusK}), as well as a particular treatment of the solar neutrino
fluxes. The dashed contour that extends to the smallest values of $\tan^2\theta_{12}$ corresponds to
treating the $^7$Be and all other solar neutrino fluxes as free parameters, while the contour that
extends to the largest values of $\tan^2\theta_{12}$ was computed by assuming that all of the solar
neutrino fluxes except the $^8$B neutrino flux are constrained by the estimated values and uncertainties
obtained from the standard solar model~\cite{bp00}. A 5\% measurement of the $\nu-e$ scattering rate for
$^7$Be solar neutrinos produces the solid intermediate curve in figure~\ref{fig:be7fig}. The contours
were calculated by marginalizing over $\theta_{13}$, $f_{\rm B}$, and $f_{\rm Be}$. \label{fig:be7fig}}}

The solid curve in figure~\ref{fig:be7fig} represents the expected $1\sigma$ allowed region after a 5\%
measurement of $\left[{\rm ^7Be}\right]_{\rm \nu-e}$. This 'post-$^7$Be' contour is not significantly
smaller than the two other contours in figure~\ref{fig:be7fig}. The two other contours represent the
result of analyzing with different options all the current solar neutrino data plus three years of
simulated KamLAND data.  The two options are to allow all the solar neutrino fluxes to be free
parameters, subject to the luminosity constraint, or to treat only the $^8$B solar neutrino flux as a
free parameter.

We have also explored the effect of a 1\% measurement of the $^7$Be neutrino-electron scattering rate on
the existing bound for the sterile fraction, $\sin^2 \eta$. We find that even a 1\% $^7$Be experiment
does not improve the existing limits derived in  section~\ref{subsec:sterile}.

We conclude that a measurement of the $^7$Be solar neutrino flux will not contribute much to the
knowledge of the allowed region in oscillation parameter space unless there is additional new physics
beyond neutrino oscillations or unless one of the previous solar neutrino experiments has a large
previously unrecognized systematic uncertainty or bias.

\subsection{How accurately will a $^7$Be neutrino experiment determine solar neutrino fluxes?}
\label{subsec:howbefluxes}

In the following subsection,  section~\ref{subsubsec:learnabout7Be}, we study what  can be learned about
the $^7$Be solar neutrino flux by measuring the $^7$Be $\nu-e$ scattering rate.  In
section~\ref{subsubsec:learnaboutpp}, we investigate how much such $^7$Be experiments will tell us about
the $p-p$ solar neutrino flux.

\subsubsection{What do we learn about the $^7$Be solar neutrino flux?}
\label{subsubsec:learnabout7Be}

If $\left[{\rm ^7Be}\right]_{\rm \nu-e}$ can be measured to an accuracy of $\pm 10$\%, then, according to
table~\ref{tab:rangeslmanycases}, the experimental knowledge of the $^7$Be solar neutrino flux will be
improved by more than a factor of four over what will be known from the solar neutrino experiments plus
three years of KamLAND reactor measurements. The experimental uncertainty in determining the $^7$Be solar
neutrino flux would then be essentially equal to the current uncertainty in the standard solar model
prediction of the $^7$Be neutrino flux. A 3\% measurement of $\left[{\rm ^7Be}\right]_{\rm \nu-e}$ would
lead to an improvement of a factor of eight in the accuracy with which the $^7$Be solar neutrino flux is
known.

We conclude from a study of table~\ref{tab:rangeslmanycases} that a measurement of the $^7$Be $\nu-e$
scattering rate will lead to an enormous improvement in the experimental knowledge of the $^7$Be solar
neutrino flux. Without a specific $^7$Be solar neutrino experiment, our empirical knowledge of the $^7$Be
solar neutrino flux will be a factor of four less precise than the current uncertainty in the standard
solar model prediction [$\pm 40$\% (experimental) versus $\pm 10$\% (standard solar model) ].

How sensitive are our quantitative conclusions to the assumed value for $\left[{\rm ^7Be}\right]_{\rm
\nu-e}$? We have answered this question by carrying out global fits in which the assumed value for
$\left[{\rm ^7Be}\right]_{\rm \nu-e}$ differs by $\pm 1\sigma$ from the current best fit. The results
obtained for the ranges of the neutrino oscillation parameters and the solar neutrino fluxes by assuming
the $1\sigma$-different values are almost identical to the results shown in
table~\ref{tab:rangeslmanycases} using our current best-prediction for $\left[{\rm ^7Be}\right]_{\rm
\nu-e}$.

\subsubsection{What do we learn about the $p-p$ solar neutrino flux?}
\label{subsubsec:learnaboutpp}

We have already seen from the discussion of table~\ref{tab:rangeSplusKppb8be7} that the $p-p$ solar
neutrino flux is already known to an accuracy of $\pm 2$\% when the luminosity constraint is imposed. By
comparing the first rows of table~\ref{tab:rangeSplusKppb8be7} and table~\ref{tab:rangeslmanycases}  that
additional measurements by KamLAND are not likely to reduce greatly the uncertainty in the experimental
measurement of the $p-p$ flux.

However, a $^7$Be measurement can enormously improve our empirical knowledge of the $p-p$ neutrino flux.
A measurement of $\left[{\rm ^7Be}\right]_{\rm \nu-e}$ to an accuracy of 5\% will improve our
experimental knowledge of the $p-p$ flux by a factor of four, to an extraordinary $\pm 0.5$\%. A 3\%
would reduce the experimental uncertainty to $\pm 0.4$\% , a factor of five improvement. No significant
further improvement would be achieved by a 1\% measurement.

We investigate  in section~\ref{subsec:cnoluminositybe} what constraints can be placed on the
CNO-generated luminosity of the Sun, and in section~\ref{subsec:totalluminositybe} on the total solar
luminosity,  from a $^7$Be neutrino-electron scattering experiment.

\subsubsection{What do we learn about the CNO solar neutrino fluxes?}
\label{subsubsec:learnaboutcnofluxes}

A measurement of the $^7$Be solar neutrino flux would significantly improve
 our current very poor experimental knowledge of the CNO neutrino fluxes (cf. equation~\ref{eq:o15pluslum}).
 Assuming a measurement of the $^7$Be neutrino flux accurate to 5\% and three more years of operation
 of KamLAND, we find that the $^{15}$O flux (in units of the standard solar model prediction~\cite{bp00})
 could be constrained to be

\begin{equation}
f_{\rm O}=0.0^{+0.5}_{-0.0}(^{+3.0}_{-0.0})\, . \label{eq:o15pluslum7be}
\end{equation}
Thus a $\pm 5$\% measurement of the $^7$Be solar neutrino flux  could improve the current $1\sigma$
($3\sigma$) constraint on the $^{15}$O solar neutrino flux by a factor of five (two) (cf.
equation~\ref{eq:o15pluslum}). At $1\sigma$, the $^{13}$N solar neutrino flux could be as large as $6$
times the standard solar model prediction and the $^{17}$F neutrino flux could be as large as 40 times
the solar model prediction.

A measurement of the $p-p$ solar neutrino flux would not by itself significantly improve our knowledge of
the CNO fluxes.  For example, adding the data from a simulated $\pm 5$\% measurement of the $p-p$
neutrino flux would improve the $1\sigma$ upper limit given in equation~\ref{eq:o15pluslum7be} from 0.5
to 0.4 and the $3\sigma$ limit from 3.0 to 2.8 .

\subsection{What do we learn about the CNO luminosity?} \label{subsec:cnoluminositybe}

A measurement of the $^7$Be neutrino-electron scattering rate to an accuracy of 5\% ($1\sigma$) will
constrain the CNO luminosity to
\begin{equation}
\frac{L_{\rm CNO}}{L_\odot} ~=~ 0.0_{-0.0}^{+1.1}\,\,\% ~(0.0_{-0.0}^{+3.9}\,\,\%)~~{\rm at ~1\sigma
\,(3\sigma)} ~~{(\rm with~ 5\% ~^7Be~ measurement)}\,. \label{eq:cnolimitbe}
\end{equation}
The result shown in equation~\ref{eq:cnolimitbe} is almost a factor of two improvement over the
constraint that is obtained without a direct $^7$Be measurement (cf. equation~\ref{eq:cnolimitlc} and
equation~\ref{eq:cnolimitnolc}).  The constraint on the CNO luminosity will not be significantly improved
by making a $^7$Be measurement that is more accurate than 5\%.

Whether or not the luminosity constraint is included does not affect, to the accuracy shown, the result
given in equation~\ref{eq:cnolimitbe} .

\subsection{How well can we constraint the total solar luminosity with a \boldmath$^7$Be neutrino experiment?}
\label{subsec:totalluminositybe}

Following the reasoning outlined in section~\ref{subsec:totalluminosity}, we can calculate the
constraints on the total from neutrino experiments, assuming a measurement accurate to 5\% of the $^7$Be
neutrino-electron scattering rate. We find

\begin{equation}
 \frac{L_\odot{\rm
  (neutrino-inferred)}}{L_\odot}~=~1.07^{+0.13}_{-0.13} \left(^{+0.40}_{-0.39}\right)
  ~~(5\% ~{\rm ~ ^7Be ~ experiment)}\, .
 \label{eq:lnuoverlphotonbe}
\end{equation}
The result given in equation~\ref{eq:lnuoverlphotonbe} is not changed, to the accuracy shown, if we
assume either a 3\% or a 1\% measurement of the $^7$Be neutrino-electron scattering rate.

The $^7$Be measurement as represented in equation~\ref{eq:lnuoverlphotonbe} would improve by about 50\%
the determination with neutrinos of the total solar luminosity (cf. equation~\ref{eq:lnuoverlphoton}).

\subsection{Will we learn more or less from  a CC $^7$Be experiment?} \label{subsec:cc7Be}

We have concentrated in this section on  $\nu-e$ scattering experiments since the
BOREXINO~\cite{borexino} and KamLAND~\cite{kamlandfirstpaper} detectors are already well advanced toward
measuring the $^7$Be neutrino-electron scattering rate. Moreover, there may be significant uncertainties
in the CC cross section that are not present for the $\nu_e$ scattering measurement.

 However, we note that a measurement with a CC detector with the same total experimental precision
 as for a $\nu-e$ scattering experiment would give
comparable, actually somewhat better, precision. For example, for a common measuring error of a
measurement of $\pm 5$\% ($1\sigma$), the $^7$Be CC absorption rate would yield a determination of the
total $^7$Be flux with an error ($\pm 4.5$\%, $1\sigma$) that is about three-quarters of the error ($\pm
6$\%) on the total $^7$Be neutrino flux that can be inferred from a $\nu-e$ scattering experiment. In
both cases, the global analysis includes a fit to all the available solar neutrino plus KamLAND data.

\section{What will we learn from a $\mathbf{p-p}$ solar neutrino experiment?}
\label{sec:whatpp}

To the best of our knowledge, Nakahata~\cite{nakahata} first discussed in a systematic way the
possibility of using a measurement of the  $\nu-e$ scattering rate of $p-p$ solar neutrinos to determine
more accurately the mixing angle represented by $\tan^2 \theta_{12}$. In his discussion, Nakahata assumed
that the total $p-p$ neutrino flux is given by the standard solar model with its associated uncertainties
($\pm 1$\% for the predicted $p-p$ neutrino flux) and therefore he did not apply the luminosity
constraint.

In what follows, it is convenient to use  the reduced  $\nu-e$ scattering rate [or CC (absorption) rate]
for $p-p$ neutrinos which is defined by the relation:

\begin{equation}
[p-p] ~\equiv~ \frac{\rm Observed ~ rate}{\rm BP00~ predicted ~rate}~, \label{eq:ppreducedrate}
\end{equation}
where the denominator of equation~\ref{eq:be7reducedrate} is the rate calculated with the BP00 $p-p$
neutrino flux  assuming no neutrino oscillations.

We assume throughout this section that the $^7$Be neutrino scattering rate has been measured to an
accuracy of $\pm 5$\% and that the data are available for three years of operation of the KamLAND reactor
experiment. From our current perspective, it seems likely that both the KamLAND reactor data and the
$^7$Be solar neutrino data will be available before the completion of a $p-p$ solar neutrino experiment.

 In section~\ref{subsec:fixedpp}, we assume that the total $p-p$ neutrino flux, and the uncertainty in
 the flux, are calculated
accurately using the standard solar model. Following Nakahata, we explore what can be learned from a
precise measurement of the rate, $[p-p]_{\nu-e}$, of neutrino-electron scattering of $p-p$ solar neutrinos
if we have faith in the standard solar model predictions.

 We consider a more rigorous and informative treatment in section~\ref{subsec:freepp}. In this analysis,
 the $p-p$, $^7$Be, $^8$B, and CNO neutrinos are treated as free parameters in a global
analysis of all the data, with the data including simulated measurements of the $p-p$ and $^7$Be
neutrino-electron scattering rates. We apply the luminosity constraint in the free-flux analysis
described in section~\ref{subsec:freepp}, obtaining simulated powerful limits on the $p-p$ neutrino flux
and the range of $\tan^2 \theta_{12}$.

We shall see that a measurement of $[p-p]_{\nu-e}$ to an accuracy better than $\pm 3$\% is required in
order to significantly improve our experimental knowledge of $\tan^2 \theta_{12}$.  The main reason for
the required high precision in the measurement of $[p-p]_{\nu-e}$ is that existing experiments plus the
luminosity constraint already determine the total $p-p$ neutrino flux to $\pm 2$\% (see the last column
of table~\ref{tab:rangeSplusKppb8be7}).

We show in section~\ref{subsec:totalluminositypp} that a measurement of the $p-p$ neutrino-electron
scattering rate will provide an accurate measurement of the total luminosity of the Sun from neutrino
measurements alone.

In section~\ref{subsec:pep}, we present for $pep$ neutrinos the current predictions for the rate of
$\nu-e$ scattering and compare the power of measuring the $pep$ scattering rate with the power of
measuring the $p-p$ scattering rate.

\subsection{Assuming the standard solar model $p-p$ neutrino flux}
\label{subsec:fixedpp}

\TABLE[h!t]{\caption{\label{tab:fixedpp}{\bf The effect of a measurement of the $p-p$ $\nu-e$ scattering
rate: Assume standard solar model $p-p$ neutrino flux. } The table presents, in rows two-four, the
best-fit and the $1\sigma$ ($3\sigma$) ranges computed by global fits that include a simulated
measurement of the expected $\nu-e$ scattering rate of $p-p$ solar neutrinos. The $p-p$ neutrino
scattering rate is presumed to be measured with either 5\%, 3\% or 1\% precision.  In all cases, we have
also included a simulated measurement with 5\% precision of the $\nu-e$ scattering rate of $^7$Be
neutrinos, a simulated result of three years of KamLAND data (K), and all available data from existing
solar neutrino experiments (S)~~\cite{chlorine,sage02,gallex,superk,snoccnc,snodaynight,gno,snosalt}. The
results from a global analysis that was made  using just the solar, KamLAND, and $^7$Be data are
presented in the first row of the table. The $p-p$ neutrino flux is constrained to have the same
best-estimate value and uncertainty as in the BP00 solar model prediction~\cite{bp00}. The $^8$B and
$^7$Be fluxes were treated as free parameters, but the luminosity constraint is not imposed. In
constructing each column of the table, we have marginalized over all variables except the one whose range
is shown.}
\begin{tabular}{@{\extracolsep{-3pt}}lcccc}
\hline \hline \noalign{\smallskip} Experiments & $\Delta m^2 (10^{-5} eV^2)$
&$\tan^2\theta_{12}$&$f_B$& $f_{Be}$ \\
\hline \hline \noalign{\smallskip}\noalign{\smallskip} S + K 3 yr +$^7$Be & $7.3^{+0.2}_{-0.2}$
($^{+0.9}_{-0.6}$) & $0.41^{+0.04}_{-0.04}$ ($^{+0.15}_{-0.09}$) & $1.01^{+0.04}_{-0.04}$
($^{+0.13}_{-0.13}$) & $0.98^{+0.06}_{-0.06}$ ($^{+0.19}_{-0.18}$)\\
\hline \noalign{\smallskip}\noalign{\smallskip} $+ [p-p]_{\nu-e}$ 5\% & $7.3^{+0.2}_{-0.2}$
($^{+0.9}_{-0.6}$) & $0.41^{+0.04}_{-0.04}$ ($^{+0.14}_{-0.09}$) & $1.01^{+0.04}_{-0.04}$
($^{+0.13}_{-0.12}$) & $0.98^{+0.06}_{-0.06}$ ($^{+0.19}_{-0.18}$)\\
\noalign{\smallskip}\noalign{\smallskip} $+ [p-p]_{\nu-e}$ 3\% & $7.3^{+0.2}_{-0.2}$ ($^{+0.7}_{-0.6}$) &
$0.42^{+0.03}_{-0.03}$ ($^{+0.12}_{-0.11}$) & $1.00^{+0.03}_{-0.03}$ ($^{+0.10}_{-0.11}$)
& $0.98^{+0.06}_{-0.05}$ ($^{+0.19}_{-0.18}$)\\
\noalign{\smallskip}\noalign{\smallskip} $+ [p-p]_{\nu-e}$ 1\% & $7.3^{+0.2}_{-0.2}$ ($^{+0.7}_{-0.6}$) &
$0.42^{+0.02}_{-0.03}$ ($^{+0.07}_{-0.08}$) & $0.99^{+0.02}_{-0.02}$ ($^{+0.07}_{-0.07}$)
& $0.99^{+0.05}_{-0.06}$ ($^{+0.18}_{-0.18}$)\\
\noalign{\smallskip} \hline
\end{tabular}
}

\FIGURE[!ht]{ \centerline{\psfig{figure=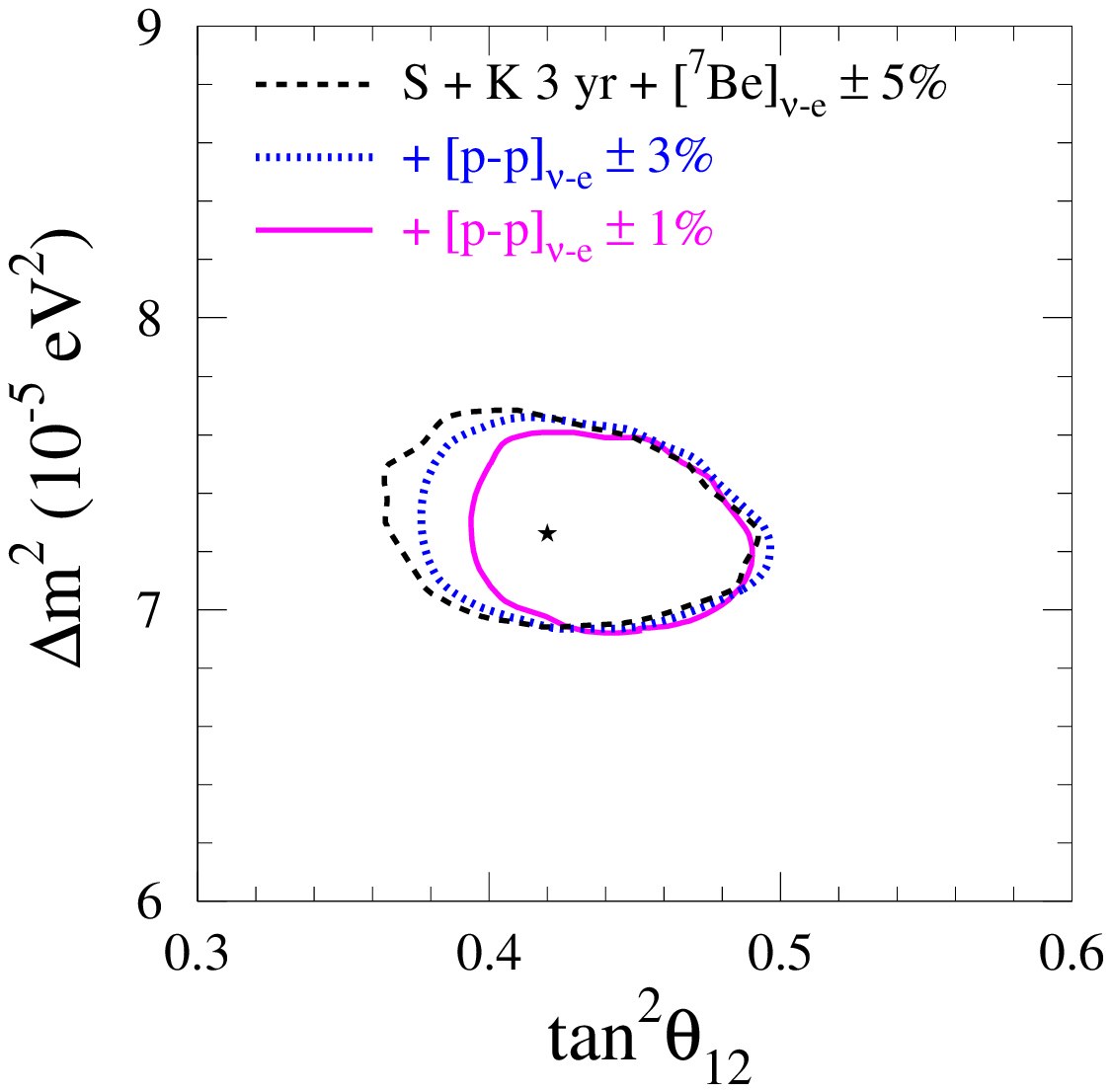,width=3.5in}} \caption{{\bf Allowed oscillation
parameters: $p-p$ + $^7$Be solar neutrino experiments plus existing Solar neutrino experiments plus 3
years of KamLAND.} The figure illustrates the effect of a future 3\% ($1\sigma$) or 1\% ($1\sigma$)
measurement of the rate, $\left[{p-p}\right]_{\rm \nu-e}$, of $p-p$ neutrino-electron scattering.  The
$p-p$, $^7$Be, $^8$B, and CNO neutrinos are treated as free parameters subject to the luminosity
constraint~\cite{luminosity}. Each contour is calculated making use of all the currently available solar
neutrino data~~\cite{chlorine,sage02,gallex,superk,snoccnc,snodaynight,gno,snosalt} plus the anticipated
data from three years of operation of the KamLAND reactor experiment (cf. figure~\ref{fig:SplusK}) plus a
5\% measurement of the $^7$Be solar neutrino flux (cf. figure~\ref{fig:be7fig}). \label{fig:ppfig}}}

Table~\ref{tab:fixedpp} shows what we can infer if we assume that the total $p-p$ flux is known to an
accuracy of $\pm 1$\% from the standard solar model predictions.

We show in the table the results of a global analysis if the expected $\nu-e$ reduced scattering rate of
$p-p$ neutrinos, $\left[{p-p}\right]_{\rm \nu-e}$, is measured to 5\%, 3\%, or 1\% accuracy. The
predicted rate for $\nu-e$ scattering with a 50 keV electron energy threshold is (see table~2 of
ref.~\cite{postkamland}):

\begin{equation}
\left[{p-p}\right]_{\rm \nu-e}~=~ 0.70\pm 0.02\, (^{+0.05}_{-0.04})\,.
 \label{eq:ppesprediction}
\end{equation}
We include in the global analysis that was used to construct table~\ref{tab:fixedpp} all the presently
known solar data, a simulation of three years of KamLAND measurements, and a measurement of the $\nu-e$
scattering rate of $^7$Be solar neutrinos,
$\left[{\rm ^7Be}\right]_{\rm \nu-e} = 0.66 (1 \pm 0.05)$.

There are, according to table~\ref{tab:fixedpp}, only modest improvements in our knowledge of the
neutrino oscillation parameters and the neutrino fluxes until the accuracy of the $p-p$ measurement becomes
better than $\pm 3$\%.  In other words, the measurement accuracy must be comparable to the quoted
accuracy, $\pm 1$\%, of the solar model calculation of the total $p-p$ flux in order to make a major
improvement. If $\left[{p-p}\right]_{\rm \nu-e}$ is measured to an accuracy of 1\%, then the
$3\sigma$ uncertainties in determining $\tan^2\theta_{12}$ and $f_{\rm B}$ will be reduced by a factor of
two. The uncertainties for $\Delta m^2$ and $f_{\rm Be}$ will be affected only by relatively small
amounts even if $\left[{p-p}\right]_{\rm \nu-e}$ is measured to 1\%.

\subsection{Free  neutrino fluxes}
\label{subsec:freepp}

\TABLE[h!t]{\caption{\label{tab:freepp}{\bf The effect of a measurement of the $p-p$ $\nu-e$ scattering
rate: free fluxes, luminosity constraint.}  The table presents, in rows two-four, the best-fit and the
$1\sigma$ ($3\sigma$) ranges computed by global fits that include a simulated measurement of the expected
$\nu-e$ scattering rate of $p-p$ solar neutrinos. The $p-p$ neutrino scattering rate is presumed to be
measured with either 5\%, 3\% or 1 \% precision.  In all cases, we have also included a simulated
measurement with 5 \% precision of the $\nu-e$ scattering rate of $^7$Be neutrinos, a simulated result of
three years of KamLAND data (K), and all currently available data from existing solar neutrino
experiments (S)~~\cite{chlorine,sage02,gallex,superk,snoccnc,snodaynight,gno,snosalt}. The global
analysis using just the solar, KamLAND, and $^7$Be results is presented in the first row of the table.
The $p-p$, $^7$Be, $^8$B, and CNO neutrino fluxes are all treated as free parameters subject to the
luminosity constraint. In constructing each column of the table, we have marginalized over all variables
except the one whose range is shown. The allowed range of $\Delta m^2$ is the same for all cases shown:
$\Delta m^2 =  7.3 \pm 0.2 \, (^{+0.9}_{-0.6})\, {\rm eV^2} $}
\begin{tabular}{lccc}
\hline \noalign{\smallskip} Experiments &$\tan^2\theta_{12}$&$f_{\rm Be}$& $f_{p-p}$ \\
\hline \noalign{\smallskip}\noalign{\smallskip} S + K 3 yr + $^7Be$ & $0.41^{+0.04}_{-0.04}$
($^{+0.15}_{-0.09}$) & $1.09^{+0.05}_{-0.07} (^{+0.17}_{-0.31})$
& $1.008^{+0.005}_{-0.005}$ ($^{+0.013}_{-0.024}$)\\
 \hline\noalign{\smallskip}\noalign{\smallskip} $+ [p-p]_{\nu-e} ~~\pm 5\%$
& $0.41^{+0.04}_{-0.04}$
($^{+0.15}_{-0.09}$) & $1.09^{+0.06}_{-0.06} (^{+0.18}_{-0.30})$
& $1.008^{+0.004}_{-0.006}$ ($^{+0.012}_{-0.024}$)\\
\noalign{\smallskip}\noalign{\smallskip} $+ [p-p]_{\nu-e} ~~\pm 3\%$ & $0.42^{+0.03}_{-0.04}$
($^{+0.11}_{-0.09}$) & $1.10^{+0.05}_{-0.07} (^{+0.16}_{-0.31})$
& $1.007^{+0.004}_{-0.005}$ ($^{+0.013}_{-0.023}$)\\
\noalign{\smallskip}\noalign{\smallskip} $+ [p-p]_{\nu-e} ~~\pm 1\%$ & $0.43^{+0.02}_{-0.03}$
($^{+0.07}_{-0.07}$) & $1.11^{+0.04}_{-0.08} (^{+0.15}_{-0.33})$
& $1.006^{+0.003}_{-0.005}$ ($^{+0.012}_{-0.021}$)\\
\noalign{\smallskip} \hline
\end{tabular}
}

Suppose we treat the $p-p$, $^7$Be, $^8$B, and CNO solar neutrino fluxes as  free parameters and impose the
luminosity condition. Suppose also the $p-p$ and $^7$Be neutrino-electron scattering rates are measured.

In this case, table~\ref{tab:freepp} and figure~\ref{fig:ppfig} show that a $\pm 1$\% measurement of the
rate of $\nu-e$ scattering rate by $p-p$ solar neutrinos will improve by a factor of two our experimental
knowledge of the allowed range of $\tan^2 \theta_{12}$, but will not significantly improve our knowledge
of the other solar neutrino fluxes or of $\Delta m^2$. An accuracy of better than  $\pm 3$\% is
necessary, according to table~\ref{tab:freepp}, to  decrease the allowed range of  $\tan^2 \theta_{12}$
by more than 15\%.

A 1\% measurement of the $p-p$ scattering rate will also improve the bound on the sterile neutrino
fraction, assuming that there are no sterile neutrinos.  Using the existing solar data and simulated
three years of KamLAND data plus a 5\% measurement of the $^7$Be neutrino-electron scattering rate and a
1\% measurement of the $p-p$ neutrino-electron scattering rate, we find that

\begin{equation}
\sin^2 \eta ~\leq ~ 0.08 \, .
 \label{eq:sterilefractionpp}
\end{equation}
The result shown in equation~\ref{eq:sterilefractionpp} was obtained by allowing all of the solar
neutrino fluxes to be free parameters and imposing the luminosity constraint. Thus a 1\% $p-p$
measurement can improve the limit on $\sin^2 \eta $ by about  20\% (cf. the discussion in
section~\ref{subsec:sterile}).  A 3\% measurement of the $p-p$ neutrino-electron scattering rate will not
significantly improve the limit on $\sin^2 \eta $.

What can more precise solar neutrino experiments tell us about the mixing angle $\theta_{13}$? A $p-p$
solar neutrino experiment must be more accurate than 5\% in order to improve the present limit that is
derived from reactor, atmospheric, and solar neutrino data (see figure~\ref{fig:theta13} and
ref.~\cite{3nuupdate}). A $p-p$ measurement accurate to 1\% will, when analyzed together with a 5\%
$^7$Be solar neutrino experiment, all presently available solar neutrino and reactor data,  and three
years of KamLAND data, be capable of constraining $\theta_{13}$ to $\sin^2 \theta_{13} < 0.036$ at
$3\sigma$. Figure~\ref{fig:theta13} illustrates this modest improvement compared to the current limit of
$\sin^2 \theta_{13} < 0.05$. A 1\% $^7$Be solar neutrino experiment will not improve the bound shown in
figure~\ref{fig:theta13} without an accurate $p-p$ experiment.

We quote in all of our tables in this paper the allowed ranges after marginalizing over every parameter
except the one of special interest. We have, for example, marginalized over all other quantities except
$\sin^2 \eta$ in obtaining equation~\ref{eq:sterilefractionpp}. This marginalization procedure is not the
common practice in the field of neutrino oscillations. Because what we mean by errors is somewhat
non-standard (but we think our procedure is valid), we have included in
section~\ref{subsec:marginalization} and also in section~\ref{subsec:principalresults} discussions of the
effect of marginalizing over input parameters.

\subsection{How accurately will a $p-p$ neutrino experiment determine the solar luminosity?}
\label{subsec:totalluminositypp}

 We have discussed in section~\ref{subsec:totalluminosity} and
section~\ref{subsec:totalluminositybe} how neutrino experiments can be used to measure the total energy
generation rate of the Sun.  We have carried out a global analysis of all of the existing solar data plus
a simulated 5\% measurement of the $p-p$ neutrino-electron scattering rate as well as three years of
simulated data of KamLAND and a 5\% measurement of the $^7$Be neutrino-electron scattering rate. We find

\begin{equation}
 \frac{L_\odot{\rm
  (neutrino-inferred)}}{L_\odot}~=~0.99^{+0.04}_{-0.04} \left(^{+0.15}_{-0.13}\right)
  ~~(5\% ~ p-p ~{\rm  experiment})\, .
 \label{eq:lnuoverlphotonpp5}
\end{equation}

Comparing equation~\ref{eq:lnuoverlphotonbe} and equation~\ref{eq:lnuoverlphotonpp5}, we see that a 5\%
$p-p$ measurement can improve the neutrino measurement of the solar luminosity by a factor of about three.

The constraints become even a factor of two stronger if the $p-p$ measurement is made to an accuracy of
1\%. In this case, we find

\begin{equation}
 \frac{L_\odot{\rm
  (neutrino-inferred)}}{L_\odot}~=~0.99^{+0.02}_{-0.02} \left(^{+0.07}_{-0.06}\right)
  ~~(1\% ~ p-p ~{\rm  experiment})\, .
 \label{eq:lnuoverlphotonpp1}
\end{equation}

A $p-p$ measurement to an accuracy of 1\% would make a truly fundamental contribution to our knowledge of
stellar energy generation. The neutrino measurements could be used to bound to $\pm 2$\% the
contributions of all sources of energy other than the low-energy fusion reactions among light elements
(the well-known $p-p$ chain and CNO cycle).

Unfortunately,  a measurement of the $p-p$ neutrino electron scattering rate to an accuracy of 1\% will
not improve the constraints on the fraction of the solar luminosity that is produced by CNO fusion
reactions. That is to say, the limits summarized in equation~\ref{eq:cnolimitbe} are not tightened, to
the quoted accuracy, by a measurement  of the $p-p$ scattering rate to a precision of 1\%.

\subsection{What about the $pep$ line?}
\label{subsec:pep}

In this subsection, we explore what can be learned by studying the $pep$ solar neutrino flux. The
interested reader can see in ref.~\cite{pep7be} an early discussion of the theoretical questions that can
be addressed by combining the $pep$ measurement with other solar neutrino measurements. Moreover, that
has recently been renewed interest in the experimental possibilities for measuring the $pep$
neutrino-electron scattering rate at a deep underground site~\cite{mark}.

We present in section~\ref{subsubsec:pepcurrent} the current best-estimate for the rate at which $pep$
neutrinos are scattered by electrons. In section~\ref{subsubsec:better}, we compare the power of a $pep$
measurement with the power of a $p-p$ measurement.

 The ratio of the
$pep$ to the $p-p$ neutrino flux is robustly determined by the standard solar model calculations.  The
ratio is determined more accurately than the individual fluxes because the ratio only depends weakly on
the solar model characteristics. We have therefore used the BP00 value for the ratio,
\begin{equation}
{\phi({p-p}) \over \phi({pep}) } ~=~ 425.9\, , \label{eq:ratiopeppp}
\end{equation}
in the calculations reported in this subsection (and in section~\ref{subsec:fixedpp} and
section~\ref{subsec:freepp}). However, we have verified that none of our quantitative conclusions are
affected significantly if we change the ratio given in equation~\ref{eq:ratiopeppp} by a relatively large
amount, $\pm 10$\%.

\subsubsection{The current predictions} \label{subsubsec:pepcurrent}

What is the current best-estimate prediction for the rate of scattering of $pep$ solar neutrinos by
electrons? To answer this question, we have repeated the global analysis of the existing solar and
KamLAND data as described in section~\ref{subsec:be7implications}. But in the present discussion we
present the predictions for the $pep$, $1.4$ MeV, neutrino line instead of the $^7$Be neutrino line.

If we treat the $^8$B solar neutrino flux as a free parameter, but assume that all the other neutrino
fluxes are as predicted  by the BP00 standard solar model calculation~\cite{bp00}, then the  current
best-estimate prediction for the reduced $pep$ reduced reaction rate is

\begin{equation}
\left[{pep}\right]_{\rm \nu-e}~=~ 0.65 \pm 0.02 (^{+0.06}_{-0.04})  \, .
 \label{eq:pepesprediction}
\end{equation}
If instead we treat all the neutrino fluxes as free parameters and impose the luminosity constraint, then
the predicted $pep$ neutrino scattering scattering rate is

\begin{equation}
\left[{pep}\right]_{\rm \nu-e}~=~ 0.66^{+0.02}_{-0.03} (^{+0.14}_{-0.07}) \, {\rm [all~free]}\,.
 \label{eq:freepepesprediction}
\end{equation}

\subsubsection{Which is better, $p-p$ or $pep$?} \label{subsubsec:better}

Assuming that the $pep$ neutrino flux is measured instead of the $p-p$ neutrino flux, we have repeated
the global analyses of existing and future solar and KamLAND data that are described in
section~\ref{subsec:fixedpp} and section~\ref{subsec:freepp}. We assumed for these calculations the same
accuracy for the measurements of the $pep$ neutrino flux as was assumed in section~\ref{subsec:fixedpp}
and section~\ref{subsec:freepp} for the measurements of the $p-p$ neutrino flux.

 Our global analyses show that a
measurement of the $\nu-e$ scattering rate by $pep$ solar neutrinos would yield essentially equivalent
information about neutrino oscillation parameters and solar neutrino fluxes as a  measurement of the
$\nu-e$ scattering rate by $p-p$ solar neutrinos. The estimated best-estimates and uncertainties in the
parameters are almost identical for the analyses we have carried out for $p-p$ and $pep$ neutrinos.

\section{Maximal Mixing?}
\label{sec:maximalmixing}

Is the currently allowed oscillation region for solar neutrinos consistent with maximal mixing, i.e.,
with $\tan^2\theta_{12} = 1$? And, if not, at what level is maximal mixing excluded? These are important
questions for constructing particle physics models (see, e.g.,
~\cite{conchayossi,Nir:2000xn,Altarelli:2002hx,Mohapatra:2002kn}).

We determine in section~\ref{subsec:solarpluskoday} at what level maximal mixing is excluded by using all
the currently available solar plus KamLAND data. In section~\ref{subsec:snohowwell}, we investigate the
extent to which the SNO salt-phase data~\cite{snoplans,snoestimates} will strengthen or weaken the
conclusion that exact Maximal Mixing is disfavored by solar and reactor data. In order to illustrate the
sensitivity (or insensitivity) of the results to individual experiments, we present the results of the
global analyses with and without including the chlorine experiment.

\subsection{Solar plus KamLAND data today}
\label{subsec:solarpluskoday}

We see from figure~\ref{fig:SplusK} that the $1\sigma$ allowed contour in the $\Delta m^2$-$\tan^2\theta_{12}$
plane satisfies $\tan^2 \theta_{12} < 0.52$. If we perform a global solution including all solar and KamLAND
data and marginalize over $\Delta m^2$, then maximal mixing is excluded at  $5.3\sigma$, i.e.,

\begin{equation}
\tan^2 \theta_{12} ~<~ 1~{\rm at~} 5.3\sigma .
\label{eq:maximalexcluded}
\end{equation}
This result given in equation~\ref{eq:maximalexcluded} is in good agreement with the value of $5.4\sigma$
for the extent of the exclusion that was obtained~\cite{snosalt} by the SNO collaboration using
independent computer codes.

 The result stated in equation~\ref{eq:maximalexcluded} is slightly weakened if we exclude the chlorine experiment in the global
solution. Then

\begin{equation}
\tan^2 \theta_{12} ~<~ 1~{\rm at~} 4.8\sigma ~{\rm \left[without ~ chlorine.\right]}
\label{eq:noclmaximalexcluded}
\end{equation}
If we had not marginalized over $\Delta m^2$, but instead had used the two-dimensional contours of
$\Delta m^2$-$\tan^2 \theta_{12}$ (as is sometimes done in the literature), then the results would have been
weaker than the constraints given in equation~\ref{eq:maximalexcluded} and in
equation~\ref{eq:noclmaximalexcluded}. We would have obtained that, for 2 d.o.f. , maximal mixing was
excluded at  $4.9\sigma$ (including chlorine) or  $4.4\sigma$ (excluding chlorine).

\subsection{How well did the simulations work?}
\label{subsec:snohowwell}

We described in section~\ref{subsec:simulationsno} of hep-ph/0305159, v2. how we  simulated, prior to
their announcement, the results of the SNO measurements in the salt phase (see also the footnote in
section~\ref{subsec:simulationsno} of the present version of our paper). Using simulated instead of real
SNO salt-phase data, we estimated that

\begin{equation}
\tan^2 \theta_{12} ~<~ 1~{\rm at~} 5.5\sigma ~{\rm \left[with~ simulated ~SNO ~salt-phase ~data\right]}.
\label{eq:maximalsnoexcluded}
\end{equation}

Thus the simulations correctly predicted that the salt-phase results would strengthen the exclusion of
maximal mixing by about an additional $2\sigma$ relative to the  exclusion of $3.5\sigma$ that existed at
the time the simulations were made.

\section{Discussion}
\label{sec:discussion} In section~\ref{subsec:whylowenergy}, we answer the question of why do solar
neutrino experiments below 1 MeV. We use as the basis for our answer the calculations and discussion
presented in the present paper.

We summarize in section~\ref{subsec:principalresults} our view of what are the most important ideas and
results that are described in section~\ref{sec:mswvacuum} to section~\ref{sec:maximalmixing}. In
section~\ref{subsec:whatmean}, we state our view of what all these results mean.

\subsection{Why do solar neutrino experiments below 1 MeV?}
\label{subsec:whylowenergy}

There are three primary reasons for doing low energy solar neutrino
experiments\footnote{For earlier discussions of the reasons for doing
low energy solar neutrino energy experiments, see
ref.~\cite{johnlowenergy}, e.g., and the many important talks in the
LowNu conference series:\hfill\break \hbox{$\rm
http://cdfinfo.in2p3.fr/LowNu03/$;}\hfill\break \hbox{$\rm
http://www.mpi-hd.mpg.de/nubis/www_lownu2002/index.html;$}\hfill\break
\hbox{$\rm http://www-sk.icrr.u-tokyo.ac.jp/lownu/;$}\hfill\break and \hbox{$\rm
http://www.sns.ias.edu/\sim jnb/Meetings/Lownu/$ .}}.

First, new phenomena may be revealed at low energies ($< 1$ MeV) that are not discernible at high
energies ($> 5$ MeV). According to the currently accepted LMA oscillation solution, the basic oscillation
mechanism switches  somewhat below 1 MeV from the MSW matter-dominated oscillations that prevail at high
energies to the vacuum oscillations that dominate at low energies (More precisely, we mean by the phrase
`low energies' those energies for which $\beta < \cos 2 \theta_{12}$, see
equation~\ref{eq:betaconvenient} and figure~\ref{fig:survival}). We want to know if this transition from
matter-induced to vacuum oscillations does indeed take place. If it does, is the ratio ($\beta$)  of the
kinematic term in the Hamiltonian (i.e., $\Delta m^2/2E$) to the matter-induced term($\sqrt{2} G_{\rm F}
n_{\rm e}$) the only parameter that determines the physical processes that are observed in this energy
range? Or, could there be entirely new physical phenomena that show up only at the low energies, very
long baseline,  and great sensitivity to matter effects provided by  solar neutrino experiments (see, for
example, refs.~\cite{Berezinsky:2002fa,deHolanda:2002ma})?

Second, new solar neutrino experiments will provide accurate measurements of the fluxes of the important
$p-p$ and $^7$Be solar neutrino fluxes, which together amount to more than 98\% of the total flux of
solar neutrinos predicted by the standard solar model.  These measurements will test the solar model
predictions for the main energy-producing reactions, predictions that are more precise than for the
higher-energy neutrinos.

Remarkably, the combination of a $^7$Be solar neutrino experiment and a $p-p$ solar neutrino experiment
can determine experimentally which terminating reaction of the $p-p$ chain, $^3$He-$^3$He or $^3$He-$^4$He,
is faster in the solar interior and by how much. The ratio R of the rate of $^3$He-$^3$He reactions to
the rate of $^3$He-$^4$He reactions averaged over the Sun can be expressed in terms of the $p-p$ and $^7$Be
neutrino fluxes by the following simple relation\footnote{More precisely, $\phi(^7{\rm Be})$ should be
replaced by the sum of the $^7$Be and $^8$B neutrino fluxes in the denominator of Eq.~(\ref{eq:defnR}).}:

\begin{equation}
R ~\equiv~\frac{<^3{\rm He} + ^4{\rm He}>}{<^3{\rm He} + ^3{\rm He}>} ~=~\frac{2\phi(^7{\rm
Be})}{\phi({\rm pp})~-~\phi(^7{\rm Be})}. \label{eq:defnR}
\end{equation}

The standard solar model predicts $R = 0.174$~\cite{standardmodelhistory}.  One of the reasons why it is
so important to measure accurately the total $p-p$ and $^7$Be neutrino flux is in order to test this
detailed prediction of standard solar models.  The value of R reflects the competition between the two
primary ways of terminating the $p-p$ chain and hence is a critical probe of solar fusion.

Using only the measurements of the solar neutrino fluxes, one can determine the current rate at which
energy is being produced in the solar interior and can compare that energy generation rate with the
observed photon luminosity emitted from the solar surface. This comparison will constitute a direct and
accurate test of the fundamental idea that the Sun shines by nuclear reactions among light elements.
Moreover, the neutrino flux measurements will test directly a general result of the standard solar model,
namely, that the Sun is in a quasi-steady state in which the interior energy generation rate equals the
surface radiation rate.

Third, low-energy solar neutrino experiments, $^7$Be plus $p-p$,  will make possible a precise
measurement of the vacuum mixing angle, $\theta_{12}$,  and an improved measurement of the sterile mixing
parameter $\sin^2 \eta$ (as well as a slightly improved constraint on $\theta_{13}$) (see
section~\ref{subsec:freepp}).

The low-energy measurements will provide, in addition, essential redundancy. In this paper, we have
assumed the correctness of all solar neutrino and reactor experiments that have been performed so far or
which will be performed in the future. But, the history of science teaches us that this is a dangerous
assumption. Sometimes unrecognized systematic uncertainties can give misleading results. To be sure that
our conclusions are robust, the same quantities must be measured in different ways.

\subsection{Principal results}
\label{subsec:principalresults}

We present a list of our principal results, beginning with the results that pertain to individual solar
neutrino fluxes (section~\ref{subsubsec:fluxes}), the neutrino oscillation parameters
(section~\ref{subsubsec:parameters}), and specific new experiments
(section~\ref{subsubsec:newexperiments}), and ending with technical remarks concerning statistical
significance and marginalization (section~\ref{subsubsec:technical}).

\subsubsection{Solar neutrino fluxes}
\label{subsubsec:fluxes}

\begin{description}
\item[The \boldmath${p-p}$ solar neutrino flux.]
The existing solar and reactor experiments determine the flux of $p-p$ solar neutrinos to an accuracy of
$\pm 2$\%, provided one imposes the luminosity constraint (see below and
section~\ref{sec:luminosityexplain}). The measured value is
$1.02 \pm 0.02$ times the flux predicted by
the BP00~\cite{bp00} standard solar model (see the last two entries in the last column of
table~\ref{tab:rangeSplusKppb8be7}).

\item[The $\mathbf{^7}$Be solar neutrino flux.] The existing solar plus reactor experiments provide only loose
constraints on the $^7$Be solar neutrino flux. The $^7$Be flux is restricted at $1\sigma$  to the values
$0.93^{+0.25}_{-0.63}$, corresponding to approximately a $\pm 40$\% uncertainty (see the last row  of
table~\ref{tab:rangeSplusKppb8be7}). The $\pm 40$\% uncertainty results when one varies all the principal
solar neutrino fluxes as free parameters subject to the luminosity constraint.

\item[The $\mathbf{^8}$B solar neutrino flux.] The $^8$B solar neutrino flux is robustly determined by the existing
solar and KamLAND experiments. If one treats only  the $^8$B flux as a free parameter, then the
experimentally-determined flux divided by the standard model (BP00) prediction is  $f_{\rm B} =
1.00 \pm 0.04$ ($\pm 0.13$) (see the second row of table~\ref{tab:rangeSplusK}).
 The
uncertainty is unchanged if one allows all of the neutrino fluxes to vary as free
parameters, subject to the luminosity constraint (see the last row of table~\ref{tab:rangeSplusKppb8be7}).

\item[The CNO solar neutrino fluxes.] The existing solar neutrino experiments constrain only very poorly
the CNO fluxes. The $^{15}$O flux is the most strongly constrained by the currently available data and
even for this most favorable case the flux could be at $1\sigma$ ($3\sigma$) as large as 2.4 (6.4) times
the standard model prediction (see equation~\ref{eq:o15pluslum}). A $^7$Be experiment accurate to $\pm
5$\% could reduce the $1\sigma$ ($3\sigma$) uncertainty in the $^{15}$O neutrino flux by a factor of five
(two) (see equation~\ref{eq:o15pluslum7be}).

\item[The power of the luminosity constraint.] Table~\ref{tab:rangeSplusKppb8be7} shows that the
luminosity constraint reduces by a factor of ten the uncertainty in the experimentally-determined $p-p$
solar neutrino flux, from $\pm 0.22$ ($1\sigma$) without the luminosity constraint to $\pm 0.02$ with the
constraint. One can understand  from the approximate equation~\ref{eq:approxluminosity} why the
luminosity constraint provides such a powerful limitation on the $p-p$ flux, but is not very restrictive
for the $^7$Be solar neutrino flux.  If the real Sun is relatively close to the standard solar
model~\cite{bp00}, about 85\% of the energy generation in the model goes through the $^3$He-$^3$He
termination (producing two $p-p$ neutrinos) and only about 15\% goes through the $^3$He-$^4$He
termination (producing one $p-p$ neutrino and one $^7$Be neutrino). Therefore the condition for energy
conservation that is  expressed by the luminosity constraint limits the $p-p$ neutrino flux more tightly
than the $^7$Be neutrino flux.

The luminosity constraint does not have a significant direct effect on the allowed region for the $^8$B
neutrino flux because the $^8$B flux represents only a tiny fraction, 0.004\%,  of the energy generation
in the standard solar model. The only effect of the luminosity constraint on the allowed $^8$B range is
indirectly through the  allowed regions for $\tan^2 \theta_{12}$ and $\Delta m^2$.

\item[Comparison of neutrino luminosity with photon luminosity.]
Neutrinos produced by a particular nuclear reaction contribute an accurately known amount of energy to
the Sun. Thus an energy-weighted linear combination of the individual solar neutrino fluxes represents
the current rate at which energy is being generated in the solar interior~\cite{luminosity}.  From
neutrino measurements alone, one can measure the solar energy generation rate and then compare this
neutrino luminosity with the photon luminosity being radiated from the solar surface.  This comparison
would test the fundamental idea that nuclear fusion reactions are responsible for the energy radiated by
the Sun. Moreover, this same comparison would test a basic inference from the standard solar model,
namely, that the Sun is in a quasi-steady state in which the energy currently radiated from the solar
surface is currently balanced by the energy being produced by nuclear reactions in the solar interior.

Using existing solar and reactor data, the neutrino luminosity can be determined to about 20\% (see
equation~\ref{eq:lnuoverlphoton}). A $^7$Be solar neutrino experiment accurate to 5\% could improve this
determination to about 13\% (see equation~\ref{eq:lnuoverlphotonbe}).  The global combination of a $^7$Be
experiment, plus a $p-p$ experiment, plus the existing solar data, and three years of KamLAND would make
possible a precise determination of the solar neutrino luminosity. A $p-p$ solar neutrino experiment
accurate to 5\% would make possible a measurement of the solar neutrino luminosity to 4\% (see
equation~\ref{eq:lnuoverlphotonpp5}) and a 1\% $p-p$ experiment would determine the solar luminosity to
2\% (see equation~\ref{eq:lnuoverlphotonpp1}).
\end{description}

\subsubsection{Neutrino parameters}
\label{subsubsec:parameters}
\begin{description}
\item[The allowed region of $\mathbf{\Delta m^2}$ and \boldmath$\tan^2 \theta_{12}$.] Figure~\ref{fig:SplusK}
illustrates the currently allowed $1\sigma$ region for $\Delta m^2$ and $\tan^2 \theta_{12}$, as well as
the much smaller allowed region that is expected to result after three years of operation of the KamLAND
experiment. If only the $^8$B neutrino flux is treated as a free parameter (all other neutrino fluxes and
their uncertainties being taken from the standard solar model), then at $1\sigma$ ($3\sigma$) (see the
second row of table~\ref{tab:rangeSplusK})  $\Delta m^2 = 7.1^{+0.4}_{-0.4}$ ($^{+2.6}_{-1.6}$)and
$\tan^2 \theta_{12} = 0.42^{+0.05}_{-0.04}$ ($^{+0.21}_{-0.12}$). If instead all the principal neutrino
fluxes are treated as free variables but subject to the luminosity constraint, then (see the last row of
table~\ref{tab:rangeSplusKppb8be7})  $\Delta m^2 = 7.3^{+0.4}_{-0.6}$ ($^{+7.7}_{-2.0}$) and  $\tan^2
\theta_{12} = 0.41^{+0.05}_{-0.05}$ ($^{+0.22}_{-0.13}$). The oscillation parameters are less well constrained
if one requires that the neutrino fluxes be determined by experiment rather than adopted from the
standard solar model.

A $p-p$ solar neutrino experiment accurate to 1\% can, when combined with existing solar and reactor
data, a 5\% $^7$Be solar neutrino measurement, and three years of KamLAND data, lead to a modest
improvement in the constraint on the mixing angle $\theta_{13}$ (see figure~\ref{fig:theta13} and
section~\ref{subsec:freepp}).

\item[The active-sterile admixture.]
The sterile fraction of the incident solar neutrino flux can be described in terms of  the parameter
$\sin^2\eta$, which is defined in section~\ref{subsec:globalchi} and in  ref.~\cite{postkamland}. The
1$\sigma$ allowed range for the active-sterile admixture is  $\sin^2\eta \leq 0.10$ (see text before
equation~\ref{eq:fbsterileKfree}) if we allow all the neutrino fluxes to vary as free parameters but
impose the luminosity constraint. For the Super-Kamiokande and SNO experiments, this limit corresponds to
$f_{\rm B,\, sterile}~=~ 0.0^{+0.07}_{-0.00}$ (see equation~\ref{eq:fbsterileKfree}).

A 1\% measurement of the $p-p$ neutrino-electron scattering rate, in conjunction with improved KamLAND
measurements and a 5\% $^7$Be experiment,  can significantly improve the existing limit on $\sin^2\eta$
(see equation~\ref{eq:sterilefractionpp}) provided that there are no sterile neutrinos that show up at
low neutrino energies. The $^7$Be measurement determines the $p-p$ flux significantly accurately that one
is then sensitive to a small loss of flux to sterile neutrinos in a $p-p$ experiment. However, even a 1\%
measurement of the $^7$Be neutrino scattering rate will not, without an accurate $p-p$ measurement,
tighten significantly the bound on $\sin^2\eta$ (see discussion in section~\ref{subsec:how7be}).

\item[The vacuum-matter transition.] We define in  equation~\ref{eq:defbeta} and equation~\ref{eq:betaconvenient}
the quantity $\beta$ [$\beta= {2 \sqrt2 G_F n_e E_\nu}/{\Delta m^2}$], which governs the transition from
vacuum neutrino oscillations to matter dominated oscillations. For values of $\beta < \cos 2
\theta_{12}$, vacuum (kinematic) oscillations are dominant while for $\beta > 1$, matter (LMA)
oscillations are dominant.

 Figure~\ref{fig:survival} illustrates the two limiting regimes, vacuum and matter oscillations,
and the transition region between them.  In the Sun, the vacuum-matter transition occurs somewhere near
$1$ MeV.

One of the principal goals of future solar neutrino experiments is to try to find direct evidence for the
vacuum-matter transition.
\end{description}

\subsubsection{New experiments}
\label{subsubsec:newexperiments}
\begin{description}
\item[What additional KamLAND measurements will teach us.] We have simulated the results of a full three
years of KamLAND operations (see section~\ref{subsec:simulatedkamland}), assuming that the true result
values of the oscillation parameters are equal to the current best-fit parameters. Performing a global
analysis of the simulated KamLAND 3 yr data and all the currently available solar neutrino data, we find
that the additional KamLAND data will reduce the allowed $3\sigma$ region for $\Delta m^2$ by about a
factor of two (compare the second and third rows of  table~\ref{tab:rangeSplusK} and
table~\ref{tab:rangeSplusKb8be7}). The allowed regions for $\tan^2 \theta_{12}$ and $f_{\rm B}$ will be
reduced less dramatically, but still significant, by about a factor of one-fourth.

\item[A $\mathbf{^7}$Be solar neutrino experiment.] The predicted rate (in terms of the standard model rate)
for a $^7$Be neutrino-electron experiment is
very well determined if one assumes the standard solar model is correct:
$\left[{\rm ^7Be}\right]_{\rm
\nu-e}~=~ 0.66\pm 0.02\, ~(1\sigma)$ (see equation~\ref{eq:be7esprediction}).  However, if one does not
assume the correctness of the solar model, then the uncertainty in the prediction is about nine times
worse:  $\left[{\rm ^7Be}\right]_{\rm \nu-e}~=~ 0.58 ^{+0.14}_{-0.23} $ (see
equation~\ref{eq:be7allfreeesprediction}).

We need an experiment to measure directly the  flux of $^7$Be solar neutrinos!

How accurate does the $^7$Be experiment have to be in order to provide important new information? This
question is answered in table~\ref{tab:rangeslmanycases}.  A measurement of the $\nu-e$ scattering rate
accurate to $\pm 10$\% or better will reduce by a factor of four the uncertainty in the measured $^7$Be
neutrino flux. Moreover, the $10$\% $^7$Be flux measurement will reduce the uncertainty in the crucial
$p-p$ flux by a factor of about 2.5.  A $^7$Be measurement accurate to $\pm 3$\% would provide another
factor of two improvement in the accuracy of the $^7$Be and $p-p$ solar neutrino fluxes.

All of these improvements are measured with respect to what we expect can be achieved with three years of
operation of the KamLAND experiment (see the top row of table~\ref{tab:rangeslmanycases}), which is
likely to be completed before a $^7$Be solar neutrino experiment is completed.  Comparable information
can be obtained from a CC (neutrino absorption) experiment and from a neutrino-electron scattering
experiment if both are performed to the same accuracy.

Contrary to what some authors have stated, a $^7$Be solar neutrino experiment is not expected to provide
significantly more accurate values for the neutrino oscillation parameters than what we think will be
available after three years of operation of KamLAND (see table~\ref{tab:rangeslmanycases}).

\item[A \boldmath${p-p}$ solar neutrino experiment.] According to the standard solar model,about 91\% of the
total flux of the neutrinos from the Sun is in the form of the low energy ($<0.42$ MeV) $p-p$ neutrinos.
We cannot be sure that we have an essentially correct description of the solar interior until this
fundamental prediction is tested. Moreover, the $p-p$ neutrinos are in the range where vacuum
oscillations dominate over matter effects, so observing these low-energy neutrinos is an opportunity to
test in a crucial way also our understanding of the neutrino physics.

If we really know what we think we know, if the standard solar model is correct to the stated accuracy
($\pm 1$\% for the total $p-p$ neutrino flux) and if there is no new physics that shows up below $0.4$
MeV, then table~\ref{tab:freepp} shows that a measurement of the $p-p$ flux to an accuracy of better than
$\pm 3$\% is necessary in order to significantly improve our experimental knowledge of $\tan^2
\theta_{12}$. The main reason why such high accuracy is required is that the existing experiments, if
they are all correct to their quoted accuracy, already determine the $p-p$ solar neutrino flux to $\pm
2$\%.  (We assumed in constructing table~\ref{tab:freepp} that three years of KamLAND reactor data will
be available, as well as $\pm 5$\% measurement of the  $^7$Be neutrino-electron scattering rate. The
$^7$Be measurement does not contribute significantly to the measurement accuracy for $\tan^2 \theta_{12}$.)

As described above, an accurate measurement of the $p-p$ solar neutrino flux will provide a direct test
of the fundamental ideas underlying the standard solar model. The $p-p$ measurement will make possible
the determination of the total solar luminosity from just neutrino experiments alone. The neutrino
luminosity can be compared with the photon luminosity to check whether nuclear fusion reactions among
light elements is the only discernible source of solar energy and whether the Sun is in an approximate
steady state in which the rate of interior energy generation equals the rate at which energy is radiated
through the solar surface.

The $pep$ neutrinos (a $1.4$ MeV neutrino line) can give essentially the same information as the $p-p$
neutrinos.

\item[Maximal mixing; the SNO salt-phase data.] The
solar (without the SNO salt-phase data) plus reactor data disfavor neutrino oscillation solutions with
maximal mixing, i. e., $\tan^2 \theta_{12} = 1$, at a confidence level of $3.5\sigma$. The addition of
the most recent solar data, excludes maximal mixing at a confidence level of $5.3\sigma$ (see
equation~\ref{eq:maximalexcluded}).

\item[What if one experiment is wrong?] The implications of solar neutrino experiments for physics and
astronomy are too important to be allowed to depend upon just one experiment, no matter how well that
experiment appears to have been performed. We have therefore checked the sensitivity of some of our
results to removing the chlorine experiment from the total data set.

The principal reason for choosing the chlorine experiment to test the robustness of our conclusions is
that the experiment supplies an important number. The total capture rate of solar neutrinos by $^{37}$Cl
affects significantly some of our inferences. Although the chlorine experiment has been tested internally
in different ways and has been carried out with exemplary care and skill~\cite{chlorine}, the full
experiment has not been directly checked by any other solar neutrino experiment.

If we include the chlorine experiment in the total data set, the reduced $^7$Be neutrino flux implied by
all the data is  $f_{\rm Be} = 0.93^{+0.25}_{-0.63}$
(see table~\ref{tab:rangeSplusKppb8be7}).  If the
chlorine experiment is removed from the data set then
$f_{\rm Be} = 1.24^{+0.31}_{-1.24}$ (see
equation\ref{eq:relaxcl}). Thus removing the chlorine experiment degrades the
existing $^7$Be neutrino
flux measurement from a  $\pm 47$\% measurement to a
 $\pm 63$\% measurement.

Removing the chlorine experiment from the data set results in a relatively
 slightly weaker
conclusion regarding maximal mixing. Instead of maximal mixing being disfavored by  $5.3 \sigma$, as it
is with the full data set, maximal mixing would only be disfavored by $4.8 \sigma$ if the chlorine
experiment is omitted (cf. equation~\ref{eq:maximalexcluded} with equation~\ref{eq:noclmaximalexcluded}.)
\end{description}

\subsubsection{Statistical significance and marginalization}
\label{subsubsec:technical}
\begin{description}
\item[Statistical significance.] Whenever we present the allowed range of a particular variable [e.g.,
$\Delta m^2$, $\tan^2 \theta_{12}$, $\sin^2\theta_{13}$, $\sin^2 \eta$, $f_{\rm B}$, $f_{\rm Be}$,
$f_{p-p}$, or $L_\odot({\rm CNO})$], we have marginalized over all parameters except the quantity singled
out for special interest. This procedure is illustrated in section~\ref{subsec:survival}, where the
marginalization procedure is outlined specifically for the variable $\theta_{13}$. The marginalization
procedure has also been applied (see discussion in section~\ref{subsec:marginalization}) to other derived
quantities like $[^7{\rm Be}]_{\nu-e}$, or $[p-p]_{\nu-e}$ (or $\left[{pep}\right]_{\nu-e}$) that are
themselves functions of the fundamental input variables.

The allowed ranges of $[^7{\rm Be}]_{\nu-e}$ or $[p-p]_{\nu-e}$ calculated by the marginalization
procedure described above are different from, and smaller than (by up to about 50\%), the allowed ranges
that were usually quoted in our own previous papers. There, we determined the uncertainties for
quantities like $[^7{\rm Be}]_{\nu-e}$ or $[p-p]_{\nu-e}$ by sampling the two-dimensional parameter
space, $\Delta m^2$ and $\tan^2 \theta_{12}$, at the desired confidence level, together with the other
input parameters. The marginalization procedure described here yields a well defined range independent of
the
 number of free parameters in the analysis.
 The marginalization procedure has been used to extract derived parameters as $\rm f_{B,sterile}$
~\cite{postkamland} or $\Delta m^2 \sin 2\theta_{12}$ (first ref. in~\cite{analysispostkamland}) and
it should be used in the future when calculating individual elements of the neutrino oscillation matrix
like $\cos\theta_{13}\times\cos\theta_{12}$.

\item[Technical Details.] We describe in section~\ref{sec:technical} the method we use to calculate the
global $\chi^2$, introduce the parameter that represents the active-sterile admixture, and define the
reduced solar neutrino fluxes $f_{\rm B}$, $f_{\rm Be}$, and $f_{p-p}$. We also summary the
96 data
points (see table~\ref{tab:experimental}) that we use in our global analyses. In addition, we outline how
we create simulated data for three years of KamLAND operation and for the salt-phase measurements of SNO.
\end{description}

\subsection{What does it all mean?}
\label{subsec:whatmean}

We are agnostics on the question of whether or not we know the physics and the astronomy well enough to
extrapolate accurately to a different neutrino energy domain and to different solar neutrino sources. New
physics or astronomical processes may be relevant at neutrino energies below 1 MeV that are not important
at higher energies. We have made the extrapolation in this paper in order to have a quantitative basis
for discussing what to expect and what we might learn.

\acknowledgments We are grateful to M. C. Gonzalez-Garcia, C. Lunardini, and E. Lisi for valuable
comments, suggestions, and insights. We are indebted to S. Freedman for a stimulating conversation
regarding what can be learned from a $^7$Be solar neutrino experiment, to A. McDonald for a valuable
discussion that sparked our simulation of the SNO salt-phase data, and to M. Chen for information about
the potential feasibility of a $pep$ solar neutrino experiment. We are grateful to many friends and
colleagues who commented on the first version of this paper, which was posted on the archive before
submission for publication. We have particularly benefited from a discussion with G. Fiorentini that
sparked our computation of the solar luminosity from purely neutrino data. We have profited from valuable
comments and stimulating suggestions by R. Lanou, D. McKinsey, F. Vissani, and L. Wolfenstein. We thank
JHEP and an anonymous referee for allowing us to update our numerical calculations following the release
on September 7, 2003 of new data from the SNO, GNO, and SAGE solar neutrino collaborations. JNB and CPG
acknowledge support from NSF grant No.~PHY0070928.

\end{document}